\documentclass[%
 reprint,
superscriptaddress,
 amsmath,amssymb,showkeys,amsthm
 aps,amsfonts,
 authblk,footmisc,cite
]{revtex4-1}
\usepackage[titletoc,title]{appendix}
\usepackage{graphicx}% Include figure files
\usepackage{dcolumn}% Align table columns on decimal point
\usepackage{bm}% bold math
\usepackage{hyperref}
\usepackage{hhline}
\usepackage{array}
\usepackage{afterpage}
\usepackage{booktabs}
\usepackage[titletoc,title]{appendix}
\usepackage[normalem]{ulem}
\usepackage{subcaption,float}
\usepackage{footmisc}
\usepackage{xcolor}
\newcolumntype{M}[1]{>{\centering\arraybackslash}m{#1}}
\begin{document}

%\preprint{APS/123-QED}

\title{Fission-fusion dynamics and group-size dependent composition in heterogeneous populations}% Force line breaks with \\

\author{Gokul G. Nair}
	\affiliation{Department of Physics, Indian Institute of Science, Bengaluru, 560 012, India}
    \affiliation{Center for Applied Mathematics, Cornell University, Ithaca, NY 14853, USA}
	\email{gokulg@iisc.ac.in}
 
\author{Athmanathan Senthilnathan}%
	\affiliation{Department of Mathematics, Indian Institute of Science, Bengaluru, 560 012, India}%
    \affiliation{Department of Ecology and Evolutionary Biology, University of Tennessee, Knoxville, TN 37916, USA}

\author{Srikanth K. Iyer}%
	\affiliation{Department of Mathematics, Indian Institute of Science, Bengaluru, 560 012, India}%

\author{Vishwesha Guttal}%
	\affiliation{Centre for Ecological Sciences, Indian Institute of Science, Bengaluru, 560 012, India}%
    \email{guttal@iisc.ac.in}

\date{\today}% It is always \today, today,

\begin{abstract}
Many animal groups are heterogeneous and may even consist of individuals of different species, called mixed-species flocks. Mathematical and computational models of collective animal movement behavior, however, typically assume that groups and populations consist of identical individuals. In this paper, using the mathematical framework of the coagulation-fragmentation process, we develop and analyze a model of merge and split group dynamics, also called fission-fusion dynamics, for heterogeneous populations that contain two types (or species) of individuals. We assume that more heterogeneous groups experience higher split rates than homogeneous groups, forming two daughter groups whose compositions are drawn uniformly from all possible partitions. We analytically derive a master equation for group size and compositions and find mean-field steady-state solutions. We predict that there is a critical group size below which groups are more likely to be homogeneous and contain the abundant type/species. Despite the propensity of heterogeneous groups to split at higher rates, we find that groups are more likely to be heterogeneous but only above the critical group size. Monte-Carlo simulation of the model show excellent agreement with these analytical model results. Thus, our model makes a testable prediction that composition of flocks are group-size dependent and do not merely reflect the population level heterogeneity. We discuss the implications of our results to empirical studies on flocking systems.
\end{abstract}

\maketitle

%\tableofcontents

\section{\label{section:intro}Introduction}
Collective phenomena and self-organization are widespread in the animal kingdom~\cite{parrish1999science,camazine2003Book,sumpter2010Book,couzin2003adv}. Theory as well as empirical works suggest that these macroscopic behaviors often emerge from simple microscopic interactions among individuals~\cite{vicsek1995prl}. Much of collective behavior theory and models assume that individuals in populations are identical~\cite{vicsek1995prl,del2018PhilTransRoySoc}. Animal populations in nature, however, are rarely homogeneous. Within conspecific social groups, heterogeneity may arise from differences in age, size, or sex. Social groups may also have dominance hierarchies including differences in behavioral tendencies such as boldness and shyness~\cite{dyer2008BehavEcol,ioannou2008Oecol,michelena2010ProcRoySocB,del2018PhilTransRoySoc}. Heterogeneity also arises when individuals of different species interact to form groups, also called mixed-species flocks~\cite{morse1970EcolMono,diamond1981Nature,sridhar2009animalbeh,stensland2003mammalreview,greenberg2000book,lukoschek2000review,sridhar2018philB}. %Studies on mixed-species flocks have been instrumental in community ecology, for instance, to infer ecological interactions in bird communities based on co-occurrence of species within flocks. 
Given the wide prevalence of individual variations among grouping species, it is pertinent to investigate how heterogeneity among individuals influences macroscopic features of collective animal behavior~\cite{gueron1996jtb,nagy2010nature,biro2006currbio,romey2013ecolmodel,herbert2013procB,aplin2014procB,farine2017procB,del2018PhilTransRoySoc}. 

Most of the previous studies that incorporate heterogeneity focus on emergent properties of single groups~\cite{gueron1996jtb,couzin2002jtb,couzin2005nature,jolles2013animalbeh,romey2013ecolmodel,farine2014animalbeh,herbert2013procB,farine2017procB,del2018PhilTransRoySoc}. Computational studies show that differences among individuals in phenotypes such as mobility, local cohesion or environmental sensing ability can lead to spontaneous assortment of phenotypes within groups~\cite{couzin2002jtb}. For example, individuals with higher speed, or `leaders' who sense environmental gradients, are often at the leading edge of groups despite the absence of any communication or signalling among group members~\cite{couzin2005nature}. Furthermore, even a relatively small proportion of such leaders can facilitate consensus decision making and transfer of information within groups~\cite{couzin2005nature,conradt2005tree}. Recently, spin-based models have been used to show analytically the existence of phase-transition like behavior in the consensus decision making in heterogeneous groups~\cite{pinkoviezky2018pre}. 
%Empirical studies show support for within group phenotypic assortment and other emergent properties~\cite{krause1996fishbiology,jolles2013animalbeh}. 

Animal populations across taxa, from insects to mammals, often form a large number of groups that frequently merge (fusion) and split (fission) among themselves~\cite{couzin2009currbiol}. Microbial populations too exhibit such dynamics either because of their self-propulsion or by being driven by their environment~\cite{joshi2017PLOSCompBio,durham2012AnnRewMarSci}. Previous studies have focussed on deriving the emergence of group size distributions in such fission-fusion populations~\cite{gueron1995MathBioSci,gueron1998JMB,durrett1999JourTheorProb,niwa2003JTB,ma2011JTB}. The role of heterogeneity, which as discussed above is widely prevalent in natural systems, has not attracted much attention in the literature on fission-fusion systems~\cite{sueur2011oikos,greenberg2000book}. Evolutionary models of collective behavior predict the emergence of heterogeneity in social, navigational or cooperative traits in fission-fusion populations~\cite{guttal2010pnas,torney2010pnas,joshi2017PLOSCompBio}. In such heterogeneous populations, each group needs to be characterized by an additional property that describes the degree of heterogeneity (referred to as group composition). In the literature on mixed-species flocks, group composition patterns are in fact used to infer species-level interactions in ecological communities~\cite{sridhar2012amnat,berry2014frontier,graves1993pnas,ulrich2010ecology,sridhar2014BehavEcolSocio}. However, group compositions are highly dynamic due to the underlying fission-fusion process among groups. %Here, we investigate this dynamic in heterogeneous fission-fusion populations and discuss their biological implicationss. 

In this paper, we develop and analyze a model of fission-fusion dynamics of heterogeneous populations. Coagulation-Fragmentation processes provide an excellent mathematical framework to model such flocking dynamics~\cite{gueron1995MathBioSci,gueron1998JMB,durrett1999JourTheorProb,niwa2003JTB,ma2011JTB,majumdar2000nonequilibrium}. One such important model, proposed by Niwa~\cite{niwa2003JTB}, assumes homogeneous groups on a fixed number of discrete sites. The two most important parameters governing the group movement between sites are the split and move rates. The former determines the rate at which a group splits into two smaller groups (fission), while the latter determines the rate at which a group moves to a new site, merging with any group (fusion) present at the new site. This fission-fusion dynamic model predicts that, in populations of identical individuals, group size distribution is approximately logarithmic. These models have been successful in predicting qualitative features of empirically observed group size distributions from the field~\cite{bonabeau1999pnas,bonabeau1995PhysRevE,niwa2003JTB,ma2011JTB,griesser2011PLOSone}. In our study, we employ this framework and generalize it to account for heterogeneity among individuals. 
%Athma: "...while the latter determines merge...". while the latter determines the rate at which groups move between different sites

For simplicity, we assume that the population consists of two types of individuals (or species). Unlike homogeneous populations, here we need to keep track of group compositions in addition to the group size distribution. We incorporate the effect of heterogeneity via increased split-rate for groups of heterogeneous composition. The resulting two daughter groups are drawn randomly from all possible partitions of the parent group. We discuss alterations to these assumptions later, but these help keep the model analytically tractable while offering interesting insights on real-world heterogeneous flocks. We first derive master equations for the group sizes and composition and obtain approximate steady-state solutions in the large population limit. We also carry out Monte-Carlo simulations of the model which show considerable agreement with the analytical solution. 

Our main finding is that the composition of the flocks depends on the group size. This is despite the merge and split rates being independent of the group size. In particular, we show that there exists a critical group size below which they are more likely to be homogeneous and contain the abundant type/species. However, groups larger than the critical size are representative of the population heterogeneity. The prevalence of heterogeneous groups is surprising, given our assumption that heterogeneous groups exhibit a higher propensity to split. 
In the Discussion section, we provide a reasoning for this phenomenon. We also discuss some interesting implications of our results for current methods used to infer interspecies interactions from mixed-species flock compositions.

\section{Merge-Split model for Heterogeneous populations}\label{section:methods}
Our formulation of the problem in heterogeneous populations is based on the merge-split model for homogeneous populations, originally conceived by Niwa~\cite{niwa2003JTB} and later analyzed by Ma et al~\cite{ma2011JTB}. Our motivation for employing this framework is two-fold. First, because of its simplicity, it is an analytically tractable framework for investigating fission-fusion group dynamics~\cite{niwa2003JTB,ma2011JTB}. Secondly, despite the simplicity of many assumptions in the model, its predictions qualitatively agree with empirically observed group size distributions in various organisms. Specifically, the model predicts that group size distribution of animals may follow a heavy-tailed and skewed distribution, described by a power-law decay followed by an exponential decay (see a review of Niwa's model in Appendix~A). Indeed, several species of fish show excellent quantitative agreement with the prediction of group-size distribution~\cite{niwa2003JTB,bonabeau1995PhysRevE,griesser2011PLOSone} while many organisms like the American buffalo, spiders, and many mammalian herbivores exhibit qualitative features of heavy-tailed distributions (see Chapter 2 of~\cite{sumpter2010Book} for more discussion on Niwa's model and its empirical validity). Given these considerations, we adopt this merge-split modeling framework and generalize Niwa's model to accommodate two species. We derive master equations from the underlying stochastic process. The derivation is non-trivial and includes some assumptions and approximations. Therefore, we only present the overall approach and key steps here, while presenting the detailed algebraic steps in the Appendix~B.

\subsection{Key Assumptions}
We assume $s$ sites without geometry and a population consisting of $N_1$ type-I individuals and $N_2$ type-II individuals which can occupy these sites with total population size, $N=N_1+N_2$. Individuals of the same type are indistinguishable. A group is defined to be the set of individuals occupying the same site at any point in time. 

As in the previous model~\cite{niwa2003JTB,ma2011JTB}, groups move from their current site to a randomly chosen site at a rate $q$ that is independent of the size of the groups. If the group lands at a site that is already occupied by another group, they merge to form a larger group with size equal to the sum of the smaller groups. 

Unlike the previous model, the groups can be heterogeneous. A group with size $ n $ of which $ k $ are of type-I, referred to as the `composition' of a group, will be denoted by the ordered pair $(n,k)$. We incorporate the role of heterogeneity via the following assumption: heterogeneous groups have a higher split rate than homogeneous ones. This assumption is justified based on previous individual-based models that indeed predict that more heterogeneous groups are less stable~\cite{gueron1996jtb,del2018PhilTransRoySoc}. More specifically, we assume a group-size-independent split rate which is a function only of the proportion of each type in the group ($k/n$ and $1 - k/n$). The split rate of an $(n,k)$-group is given by:
\begin{equation}
	p(n,k)=p_0+\frac{k}{n}\left(1-\frac{k}{n}\right)\delta
	\label{eq:split_rate}
\end{equation}
In Eq~\eqref{eq:split_rate}, $ p_0 $ is the base split-rate that is experienced by homogeneous groups (i.e.~when $k=0$ or $n$). 
%% Edit made in response to minor point 1
The {\it excess split-rate parameter}, $ \delta > 0 $ determines the maximum extent to which split rates of heterogeneous groups exceed that of homogeneous ones. The function, $p(n,k)$ is concave down with respect to the proportion $k/n$, i.e.~heterogeneous groups have a higher split rate than homogeneous ones. Groups with proportion $ k/n=0.5 $ experience the maximum split rate, $p=p_0+\frac{\delta}{4}$. 

%The assumption that merge rates are independent of group compositions is a direct consequence of locality in the underlying microscopic model. Composition dependent merge rates would violate this requirement, since it would imply that individuals can detect compositions across groups, which would require a non-local interaction. 
%Gokul: This paragraph was added later. Is it necessary? %Vishu: Lets work on this later.

When groups do split, they do so uniformly at random, i.e.~every possibility that results in two daughter groups is equally probable. Hence, heterogeneous groups are more likely to split but the mechanism of the split does not favour any type of group (see {\it Discussion} for how relaxing this assumption may influence our main results). A group $(n,k)$ splits into two groups $(k_1 + k_2, k_1)$ and $(n-(k_1+k_2), k - k_1)$ such that $k_1 \sim U(0,k)$ and $k_2 \sim U(0, n-k)$, where $U(a,b)$ is the uniform distribution on the integers in the interval $[a,b]$. The random variables are sampled conditional on $0 < k_1 + k_2 < n$, which ensures that there is a split.  After splitting, the two daughter groups occupy random sites.

%When a group splits ($(n,k)\rightarrow(n_1,k_1)+(n-n_1,k-k_1)$) such that $0 < n_1 < n$), the size and composition of the resulting group is uniformly distributed, $ n_1\sim U(1,n) $ and $ k_1\sim U(0,k) $ (where $U(a,b)$ is uniform distribution on the integers in the interval $[a,b]$).
\subsection{Transition events}\label{subsec:transition}

The number of $(n,k)$-groups at time $t$, denoted by $X(n,k,t)$, is the primary random variable of interest. We derive an equation for the rate of change of expected value of this random variable, defined as $f(n,k,t):=\mathbb{E}\left[X(n,k,t)\right]$. This is done by considering all events that will lead to a change in $X(n,k,t)$ in a small time interval. All such events, along with the resulting change to the number of {\it focal groups} $(n,k)$, are listed below in Fig~\ref{fig:transitions}. There are six such events that can lead to a change in the number of focal groups--- three merge and three split events.
%%Edit in response to Minor point 2

We denote the rates of these events as $P_{\alpha}$ if it's a split event and $Q_{\alpha}$ if it's a merge event. The subscript $\alpha$ indicates the change in the total number of groups, $X(n,k,t)$. Thus,
\begin{enumerate}
\item $Q_{\alpha}(t)$: A merge event changes $X(n,k,t)$ by $\alpha \in \{-2,-1,1\}$.
\item $P_{\alpha}(t)$: A split event changes $X(n,k,t)$ by $\alpha \in \{-1,1,2\}$.
\end{enumerate}
A graphical representation of all transition events are shown in Fig~\ref{fig:transitions} and the exact expressions for these rates are derived in Appendix~B.1.

\subsection{Dynamical equations}\label{subsec:dynamical}
Using the above notations for the rates of various events, we obtain the following equation that determines how the expected number of groups of composition $(n,k)$ changes with time,
%\begin{widetext}

\begin{equation}
	\begin{aligned}
    	\frac{d f(n,k,t) }{dt}=&\mathbb{E}\left[Q_{+1}(t)\right]-\mathbb{E}\left[Q_{-1}(t)\right]-2\mathbb{E}\left[Q_{-2}(t)\right]\\
        &+\mathbb{E}\left[P_{+1}(t)\right]-\mathbb{E}\left[P_{-1}(t)\right]+2\mathbb{E}\left[P_{+2}(t)\right].
	\end{aligned}
\label{eq:het_master_eq_dependent}
\end{equation}

%\end{widetext}
\noindent
This is also known as the master equation. In the large $N$ limit it is reasonable to assume that the random variables $X(n,k,t)$ are pairwise independent. This allows us to rewrite the master equation as  (for details see Appendix~B.1)

%\begin{widetext}

\begin{equation}
	\begin{aligned}
   \frac{df(n,k,t)}{dt}=& \sum_{i=1}^{n-1} \sum_{j=(i+k-n) \vee 0}^{i \wedge k} q f(i,j,t) \frac{f(n-i,k-j,t)}{s}\\ 
   &-\boldsymbol{1}_{n\equiv 0(mod~2)} \boldsymbol{1}_{k\equiv 0(mod~2)} q\frac{f(\frac{n}{2},\frac{k}{2},t)}{s}\nonumber\\ 
   &-\boldsymbol{1}_{n\neq 1}p(n,k)f(n,k,t)\\
   &+ \sum_{i=n+1}^N \sum_{j=k \vee (i-N_2)}^{(i+k-n) \wedge N_1} \frac{2p(i,j)f(i,j,t)}{(j+1)(i-j+1)-2}\nonumber\\
   &-\frac{2q}{s}f(n,k,t)\sum_{i=1}^{N-n}\sum_{j=(n+i-k-N_2) \vee 0}^{(N_1-k) \wedge i}f(i,j,t)\\
   &+\boldsymbol{1}_{n-\frac{N_2}{2} \leq k \leq \frac{N1}{2}}2q\frac{f(n,k,t)}{s}.
   \end{aligned}
   \tag{3}\label{eq:het_master_eq}
\end{equation}

%\end{widetext}
\noindent
where we remind the reader that $p$ is the split rate, $q$ is the move rate, and $s$ is the number of sites. The notations $a \vee b$ and $a \wedge b$ represent the maximum and the minimum of $a$ and $b$, respectively. Finally, $\boldsymbol{1}$ is an indicator function defined for a statement $A$ as

%\begin{widetext}
\begin{equation}\label{eq:indicator}
\boldsymbol{1}_{A}:=
\begin{cases}
1 &\text{if } A \text{ is true}\\
0 &\text{if } A \text{ is false.}
\end{cases}\tag{4}
\end{equation}%Gokul: Indicator function was not defined earlier
%\end{widetext}
\noindent
We also write a mean-field equation by generalising the one in \cite{ma2011JTB} (Refer to Eq~(9b) in Appendix~A). The expected total number of groups, $Z(t):=\sum_{n}\sum_{k}f(n,k,t)$, obeys the following equation

%% EQUATION is spilling out

\begin{equation}
\begin{aligned}
 	\frac{dZ(t)}{dt} =& \sum_{i=2}^{N}\sum_{j=0\vee (i-N_2)}^{i\wedge N_1}p(i,j)f(i,j,t)\\
    &-\sum_{i=1}^{N}\sum_{j=0\vee(i-N_2)}^{i\wedge N_1}\Bigg( \frac{q}{s}f(i,j,t)\sum_{k=1}^{N-i}\sum_{l=(i+k-j-N_2)\vee 0}^{(N_1-j)\wedge k}\\
    &f(k,l,t)\nonumber-\boldsymbol{1}_{i\leq \frac{N}{2}}\boldsymbol{1}_{i-\frac{N_2}{2}\leq j\leq \frac{N_1}{2}}\frac{q}{s}f(i,j,t)\Bigg).
\end{aligned}
\tag{5}\label{eq:het_mean_field}
\end{equation}
\subsection{Steady-state equations}
In steady state, i.e.~$\frac{df(n,k,t)}{dt}=0$ and $\frac{dZ(t)}{dt}=0$, we derive equations relating $Z$, the expected total number of groups to $W(n,k):=\frac{f(n,k)}{Z}$, the expected proportion of $(n,k)$-groups. When the system size is large ($s\rightarrow \infty$), it is natural to assume that $Z$ also grows such that the ratio of the two, the fraction of occupied sites, also converges to a constant ($ \frac{Z}{s}\rightarrow Z_0 $). In other words, for large systems, the fluctuations in $Z$ are of a smaller order than $s$. This finally results in the following two equations:

 \begin{align*}
 0=&\sum_{i=1}^{n-1} \sum_{j=(i+k-n) \vee 0}^{i \wedge k} q W(i,j) W(n-i,k-j)\\
 &-\boldsymbol{1}_{n\neq 1}\frac{p(n,k)}{Z_0}W(n,k)\nonumber\\
 &+\sum_{i=n+1}^N \sum_{j=k \vee (i-N_2)}^{(i+k-n) \wedge N_1} \frac{2p(i,j)W(i,j)}{Z_0((j+1)(i-j+1)-2)}\\
 &-2qW(n,k),
 \tag{6}\label{eq:ss_equation}
 \end{align*}

\begin{equation}
 Z_0 = \frac{1}{q}\sum_{i=2}^{N}\sum_{j=0\vee(i-N_2)}^{i\wedge N_1}p(i,j)W(i,j,t). 
  \tag{7}\label{eq:ss_mean_field_equation}
 \end{equation}	

\noindent
Using an iterative scheme, we solve Eq~\eqref{eq:ss_equation} and Eq~\eqref{eq:ss_mean_field_equation} to obtain $W(n,k)$. A detailed description of the derivation, including all the approximations and the iterative technique is provided in Appendix~B.

\subsection{Monte-Carlo Simulations}
Using a Monte-Carlo algorithm, we simulate the system described above. We maintain a two dimensional counter, $ C(n,k) $ that stores the number of groups of size $n$ with $k$ type-I individuals. At discrete time points, a Bernoulli random variable with appropriately calculated parameter was used to decide between the occurrence of a split and merge. In the case of a split event, the group that undergoes splitting is decided using a bivariate random variable whose probability mass function, $ P(Y=(n,k)) $ is proportional to $ p(n,k)C(n,k) $, where $ p $ is the split rate. When a group splits, the number of type-I and type-II individuals in the daughter groups is uniformly distributed between 0 and the value for the parent group. Merge events are simulated in an analogous way. The initial condition for the simulation is obtained by placing $N_1$ type-I and $N_2$ type-II individuals uniformly at random on the s sites. After the system reaches steady-state we sample the counter at regular intervals to produce the distribution.

\subsection{Parameter values}
The parameters governing the dynamics of this model are the base split-rate, $p_0$, merge rate, $q$, and the excess split-rate, $\delta$. The limiting scenario, where $p_0>>q$ is uninteresting because split events dominate, and we see very few groups consisting more than a single individual. On the other hand, when $p_0<<q$, very large groups ($\sim N$) occur frequently. This is uninteresting too, since very large groups are dominated by purely combinatorial factors and thus their composition of two types will reflect that of the population. For most of our study, we fixed $p_0=1$ and chose merge rates which were of the same order of magnitude ($q=5$). This ensures that there is a sufficiently large variability in group sizes. For the same reason, we also ensured that values of excess split-rate are also in the same order of magnitude ($q=0$ to $16$). Furthermore, to ensure that assumptions of the model and analytical approximations are met, we chose large values for the population size ($N=10000$) and the number of sites ($s=10000$). %To test how sensitive our analytical approximation is to the assumption of large population and system sizes, we also considered a smaller population size ($N=1000$) with smaller system sizes ($s=1000, 500, 250$).

%Athma: After reaching the steady-state?
\subsubsection*{Data and Codes}
The source codes for the Monte-Carlo simulation and iterative solutions to Eq~\eqref{eq:ss_equation} and \eqref{eq:ss_mean_field_equation} can be found at the following link:
\url{https://github.com/nairgokul/MergeSplit}. Detailed documentation for the Monte-Carlo simulation is also provided in this repository. Data used to plot the figures can also be found here.

\section{Results}\label{section:results}
\begin{figure}[h]
	\includegraphics[width=\linewidth]{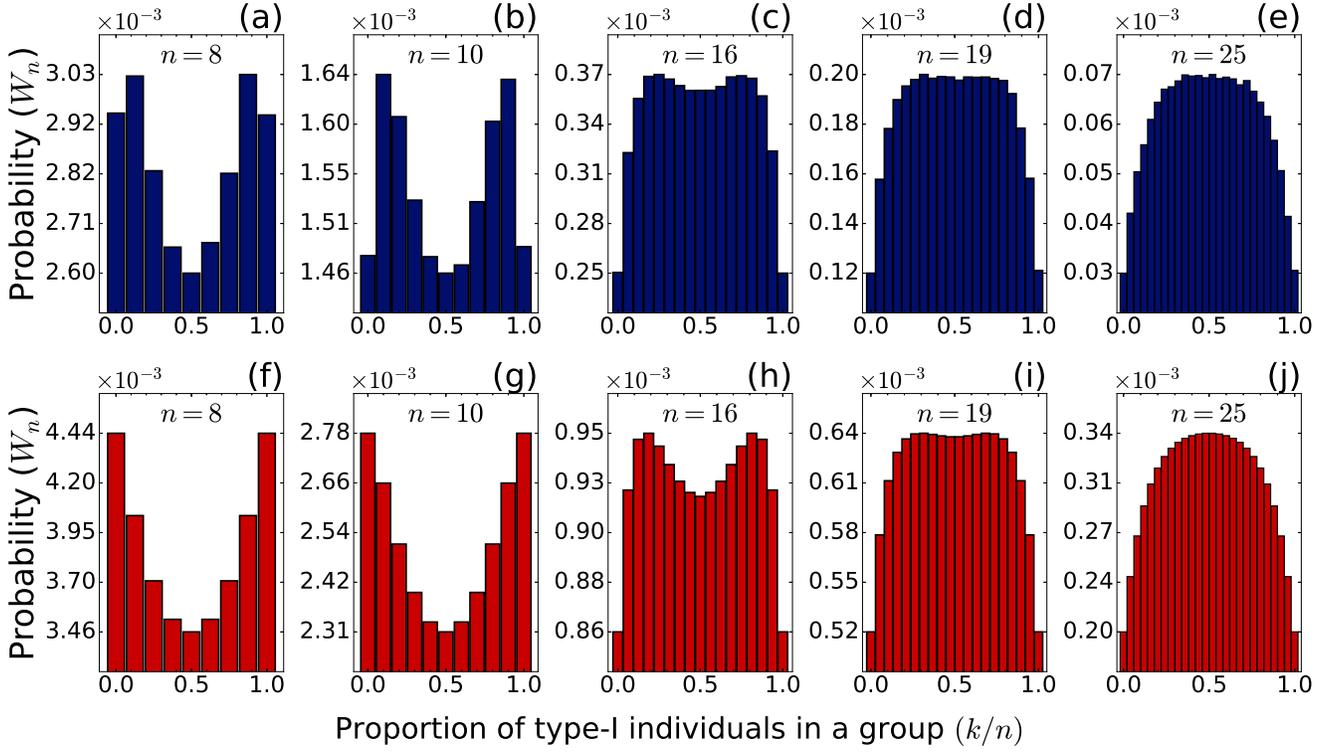}
	\caption{Flock composition depends on flock size in heterogeneous populations. We find qualitative agreement between simulation results ((a)-(e), top row) and the iterative solutions ((f)-(j); bottom row) of the analytical model equations Eq~\eqref{eq:ss_equation} and \eqref{eq:ss_mean_field_equation}. We represent flock composition for each group size $n$ by the probabilities ($W_n$; see Eq~\eqref{eq:wn}) as a function of the relative composition of type-I individuals ($k/n$). Here, we assumed that two types are equally abundant in the population ($N_1= N_2=N/2$). For small $n$, shown in (a)-(c) and (f)-(h), $W_n$ is a bimodal function with modes occurring away from the population ratio of type-I to type-II (which is 0.5); this suggests that small group sizes are dominated by one or the other type/species.  For large $n$ above a critical value, shown in (e) and (j), $W_n$ is unimodal at $k/n=0.5$, suggesting that large group sizes represent population ratio of two types. Parameters: $ s=10,000 $, $ N=10,000 $, $ N_1=5,000 $, $ p_0=1.0$, $ q=5.0 $, $ \delta=8.0 $.}
    \label{fig:cond_prob_n}
    \end{figure}

To begin, we consider populations with equal proportion of type-I individuals and type-II individuals. From Fig~\ref{fig:cond_prob_n}, we observe that the results of Monte-Carlo simulations (top row) and iterative solutions to Eq~\eqref{eq:ss_equation}-\eqref{eq:ss_mean_field_equation} (bottom row) are in qualitative agreement. We find that groups smaller than a critical size, denoted by $n_c$, are more likely to be homogeneous, that is, they are dominated by either one of type-I or type-II individuals. Groups larger than the critical size ($n_c$) are usually mixed in equal proportions. We infer this by studying the probabilities 
\begin{equation}\label{eq:wn}
W_n = \{ W(n, k); k \ge 0 \}\tag{8}
\end{equation} %% EDIT made in response to minor comment 4
of relative composition $k/n$ of type-I for various group sizes ($n$). For small groups ($n < n_c$) the distribution is bimodal with modes close to $k/n=0$ or $k/n=1$, suggesting a largely homogeneous composition of groups. As the group size increases to the critical size $n_c$, the two modes converge to a single mode at $ k/n \approx 0.5 $ (Fig~\ref{fig:cond_prob_n}), representing the greater tendency of groups to contain both types in equal proportions. The distribution remains unimodal, and thus heterogeneous, for all group sizes $n > n_c$ (Fig~\ref{fig:cond_prob_n}). This is surprising given that heterogeneous groups, for all group sizes, have a higher split rate.

\begin{figure}[h]
    \includegraphics[width=\linewidth]{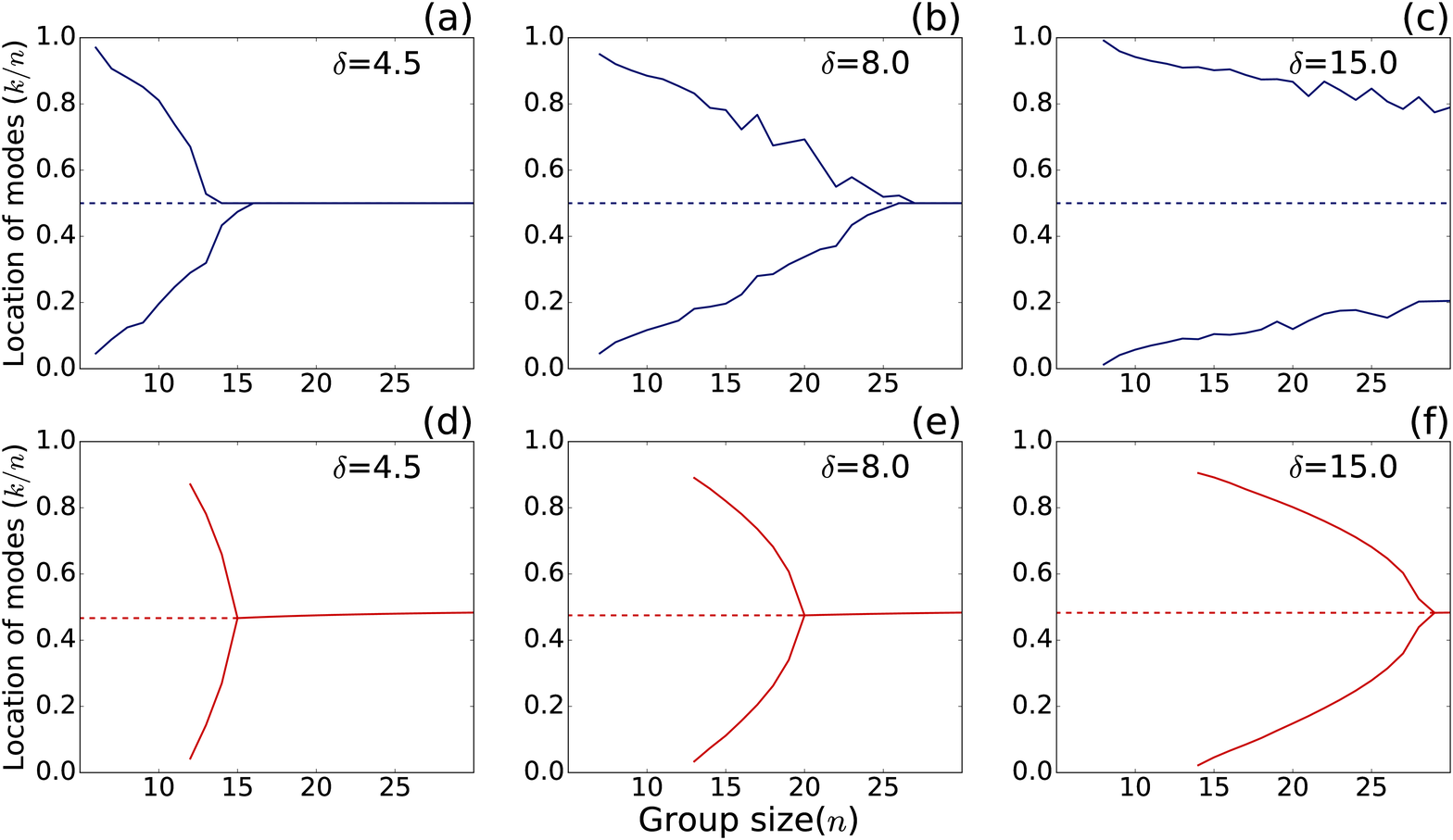}
	\captionof{figure}{Most frequently found composition for flocks, i.e.~modes of $W_n$ (see Eq~\eqref{eq:wn}), changes with group size and shows a qualitative feature similar to a pitchfork bifurcation. There is qualitative agreement between results of Monte Carlo simulations (top row; (a)-(c)) and iterative solutions of the analytical model (bottom row; d-f); specially, the agreement is even quantitatively reasonable for smaller values of excess split rate, $\delta$ ((a)-(b) and (d)-(e)) but not so for higher values of $\delta$, as seen from comparing (c) versus (f). Parameters of simulations: $s=10,000 $, $ N=10,000 $, $ N_1=5,000 $, $ p_0=1.0$, $ q=5.0 $.}
    \label{fig:bifurcation_n}
    \end{figure}

\begin{figure}[h]
   
    \includegraphics[width=\linewidth]{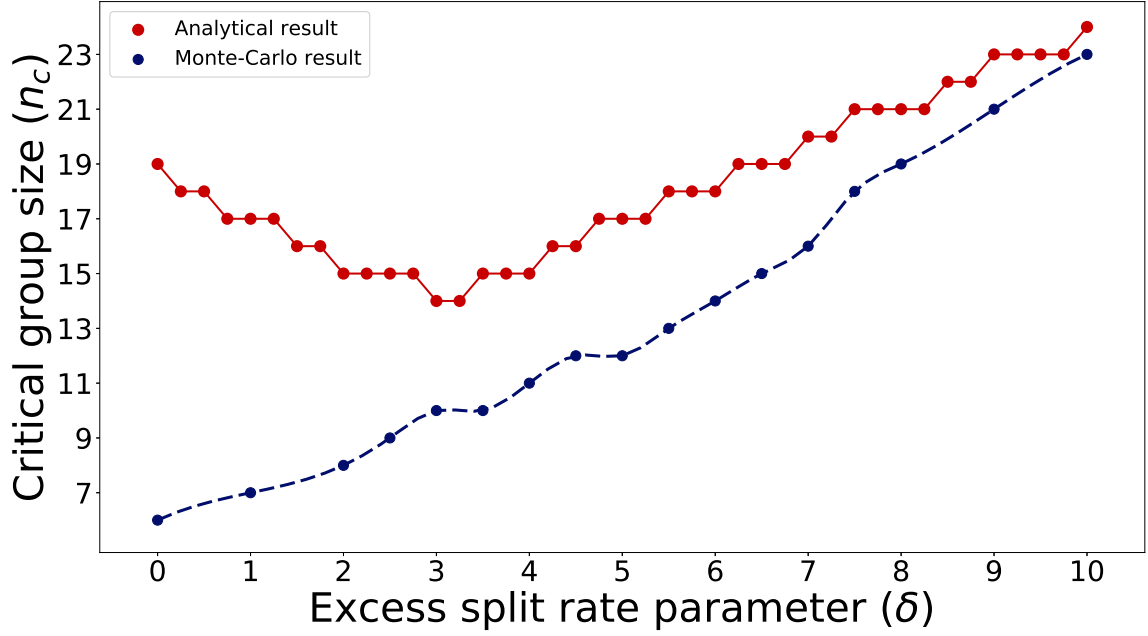}
  \caption{Dependence of $n_c$, the critical group size, on $\delta$, the excess split-rate. The results of Monte-Carlo simulations shows that $n_c$ increases approximately linearly with $\delta$. The semi-analytical result also seems to show an increasing trend for $\delta\geq 3$. However, for smaller values of $\delta$ the trend seems to be reversed. Since there is no accurate way to find the value of $n_c$ from the Monte-Carlo simulations (due to slow convergence near criticality), the blue curve must be thought of as an approximate trend line, rather than an exact one. Parameters: $s=10,000 $, $ N=10,000 $, $ N_1=5,000 $, $ p_0=1.0$, $ q=5.0 $.}
  \label{fig:deltaVSn_c}
    \end{figure}

To demonstrate the above transition from homogeneous to heterogeneous groups at a critical group size, in Fig~\ref{fig:bifurcation_n}, we plot the location of the modes of $W_n$ as a function of group size ($n$). The transition from bimodality to unimodality appears qualitatively similar to a pitchfork bifurcation~\cite{strogatz2014book}. In this bifurcation, two stable and one unstable fixed points converge to give a single stable fixed point. In our system, the modes (maxima) of the distribution $W_n$ can be viewed as stable fixed points and minima as unstable fixed points. It must be noted that the value of $n_c$ is dependent on the excess split-rate parameter ($\delta$) and increases as we increase $\delta$ (Fig~\ref{fig:deltaVSn_c}). As is evident in Fig~\ref{fig:deltaVSn_c}, both the analytical calculations and the Monte Carlo simulations predict that $n_c$ increases for larger values of $\delta$. However, for smaller values of $\delta$ ($<3$), the analytical result predicts that critical group size increases with reducing excess split-rate, which is inconsistent with Monte-Carlo simulations. We suspect that this may be related to other anomalous results observed at low $\delta$, possibly due to a violation of the assumption of independence of group compositions, a point we return to later in this section.

\begin{figure}[h]
   
    \includegraphics[width=\linewidth]{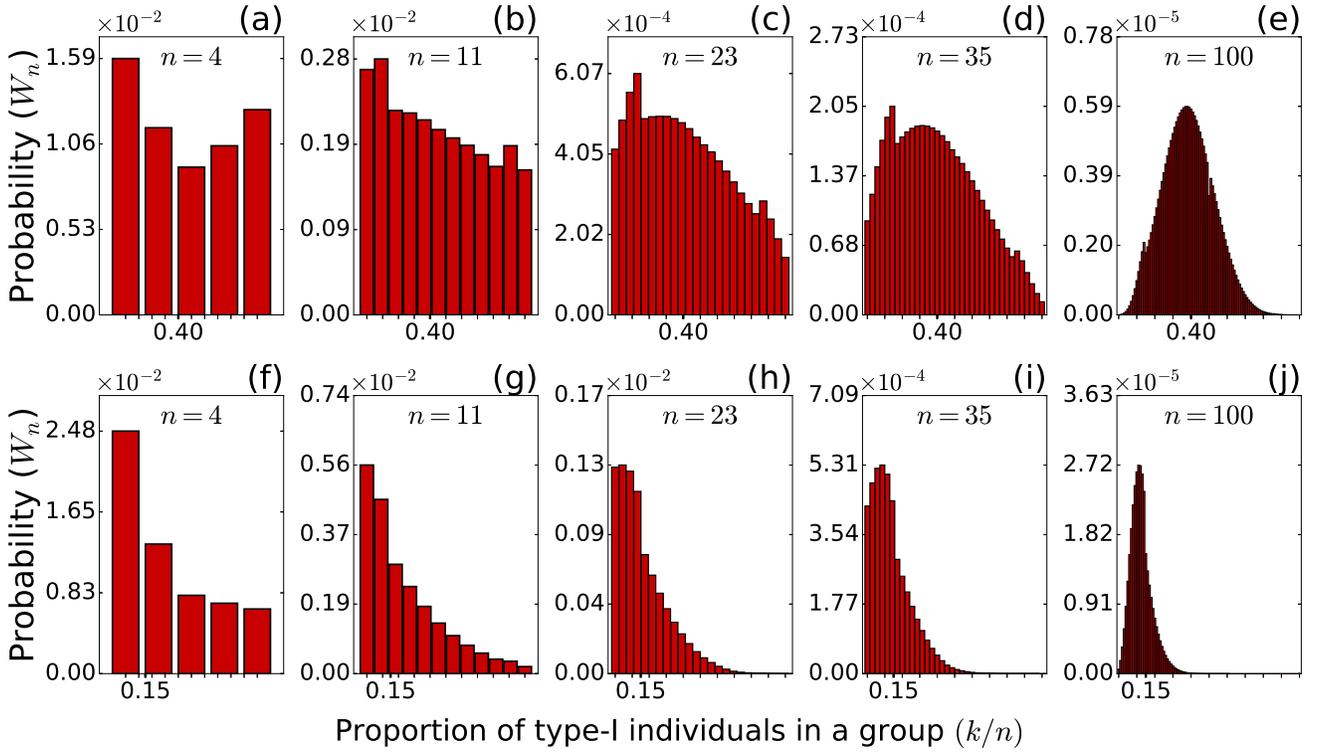}
  \caption{Flock composition for different group sizes, when the number of type-I individuals ($N_1$) is not same as that of type-II ($N_2$). These figures are obtained from the iterative solutions to the analytical model; the colour difference in the last column ($n=100$) is an artefact of high density of data. For the top row, the ratio $ N_1/N=0.40 $ and for the bottom row, $ N_1/N=0.15 $. The x-axis runs from 0 to 1 in all the panels. In the top row, we find that results are qualitatively similar to the case of the ratio being 0.5 (i.e.~Fig~\ref{fig:cond_prob_n}). In the bottom row, when the population ratio is skewed towards one type/species (bottom row), we find that the distribution is always unimodal. Nevertheless, the biological interpretation is broadly the same: smaller groups are likely to be homogeneous but also contain the abundant type/species. Larger groups, like in Fig~\ref{fig:cond_prob_n} and top row of this figure, reflect the population ratio of the two types. }
  \label{fig:skewed_population_n}
    \end{figure}

We show the plots for two cases of unequal abundances of the two types/species in the population in Fig~\ref{fig:skewed_population_n}. First, when the proportion of type-I ($N_1/N$) is closer to 0.5, we find that the above results broadly hold true (top row of Fig~\ref{fig:skewed_population_n}): As the group size ($n$) increases, the distribution $W_n$ changes from a bimodal to a unimodal distribution. Unlike the equal proportion scenario where extreme modes merge to form a unimodal distribution (Fig~\ref{fig:cond_prob_n}), in this case, the two extreme modes vanish with increasing $n$ and a mode at the population proportion emerges. Second, when the proportion of type-I ($N_1/N$) is much smaller than 0.5, the distribution remains unimodal for all group sizes. However, the mode of the distribution gradually moves from an extreme end representing homogeneous groups composed of the abundant species to one representing the population proportion ($N_1/N$). 

We remark that despite differences in the way the modes of $W_n$ behave for different population proportions of two types, our model predicts a consistent pattern of group-size dependent composition, i.e.~small group sizes are likely to be homogeneous with the abundant species whereas larger groups contain two species reflecting population proportion. These surprising qualitative features arise despite simple assumptions of the model such as group-size independent merge and split rates and an excess split-rate associated with heterogeneous groups. We provide an intuitive explanation for this in the Discussion section below.

\begin{figure}[h]
    \includegraphics[width=\linewidth]{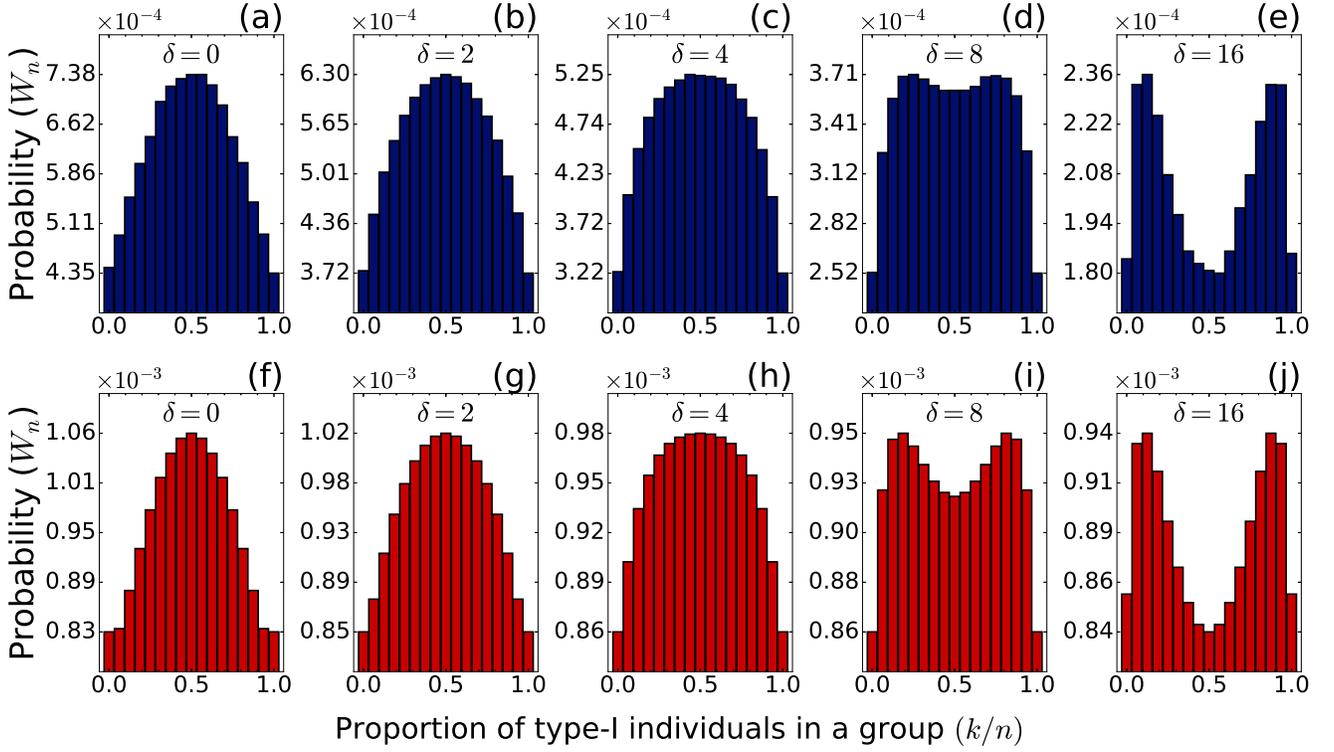}
	\caption{Flock compositions in a population where of type-I and type-II individuals appear in equal proportions. Here, we look at the compositions of groups of fixed size $n=16$ for different values of excess split-rate for heterogeneous groups ($\delta$). This shows that there is a critical split rate beyond which heterogeneous groups are less likely but group composition is bimodal. As before, (a)-(e) are results of Monte carlo simulations, while (f)-(j) are from the iterative solution of the analytical model. Parameters: $ s=10,000 $, $ N=10,000 $, $ N_1=5,000 $, $ p_0=1.0$, $ q=5.0 $, $ n=16 $.}
    \label{fig:cond_prob_delta}
    
\end{figure}

\begin{figure}[h]
	\includegraphics[width=\linewidth]{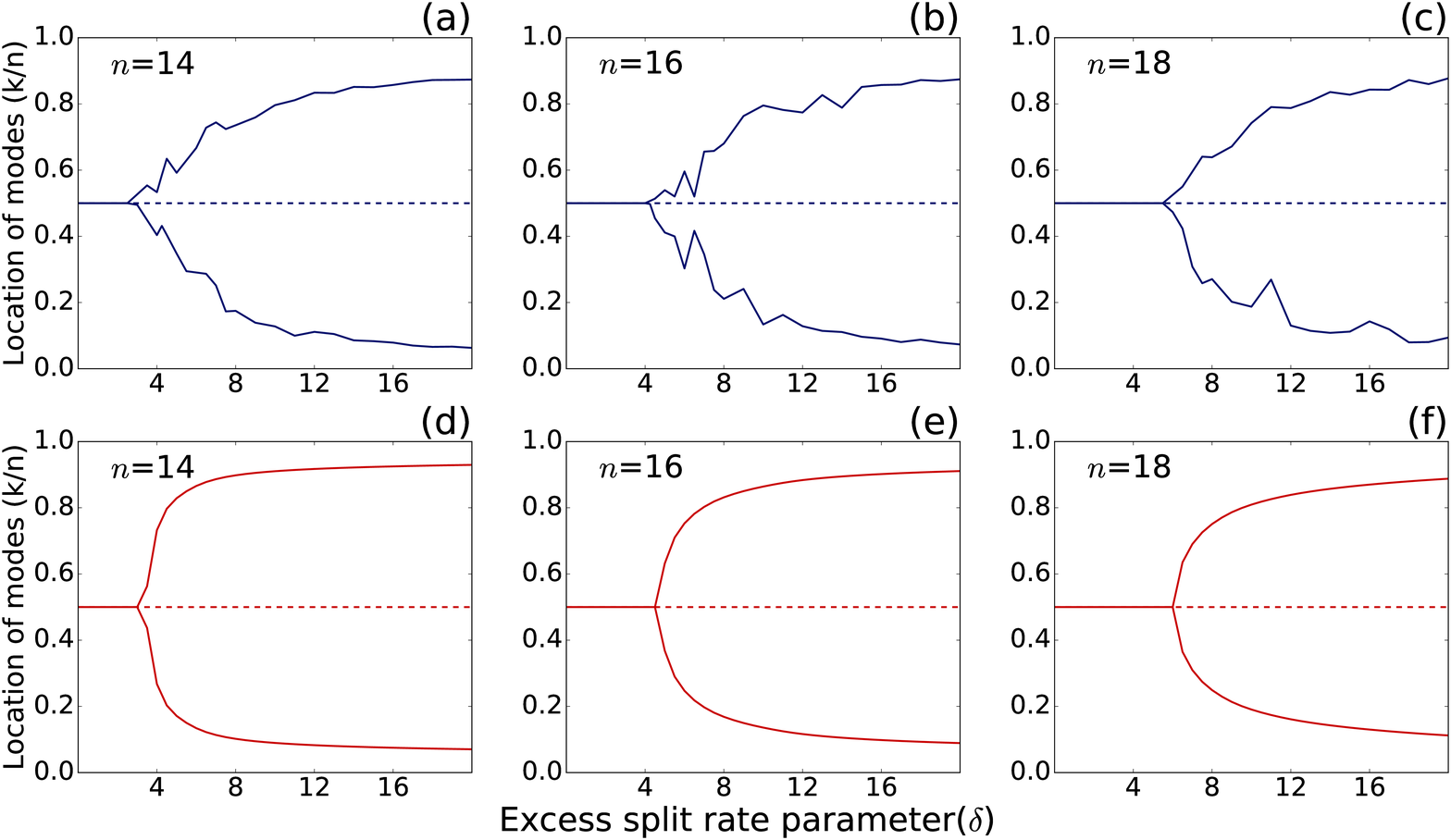}
	\caption{Most frequently found composition for flocks, i.e.~location of modes of $W_n$ (see Eq~\eqref{eq:wn}), as a function of excess split rate $\delta$, for different values of group size $n$. As before, (a)-(e) are results of Monte Carlo simulations, while (f)-(j) are from the iterative solution of the analytical model.  $ s=10,000 $, $ N=10,000 $, $ N_1=5,000 $, $ p_0=1.0$ $ q=5.0 $.}
    \label{fig:bifurcation_delta}
\end{figure}

On a similar note, we study $W_n$ as a function of the excess split-rate ($\delta$) due to group heterogeneity. We find that when $\delta$ is less than a critical value, the distribution $W_n$ has a single mode at $ k/n \approx 0.5 $ (Fig~\ref{fig:cond_prob_delta})), representing heterogeneous groups. For $\delta$ above that critical value, however, the distribution becomes bimodal with modes occurring close to $k/n=0$ and $k/n=1$, indicating higher likelihood of homogeneous groups. The location of the modes plotted as a function of excess split-rate ($\delta$) also shares qualitative features of a pitchfork bifurcation (Fig~\ref{fig:bifurcation_delta}).

We have observed that in the case where the split rate, $p_0$ is large in comparison to the merge rate, $q$, the predictions of the analysis show poor agreement with the results of Monte-Carlo simulation. In particular, when the excess split-rate parameter, $\delta$ is small, the equations predict small groups to be bimodal, but this is not the case for the results of the Monte-Carlo simulations. The results do, however, show agreement for large groups.  We also recall the inconsistency between analytical and Monte-Carlo simulations in predicting $n_c$, shown in Fig~\ref{fig:deltaVSn_c}. We suspect these are to be due to a break-down in the assumption of independence among the random variables $X(n,k)$ (see sections~\ref{subsec:transition},~\ref{subsec:dynamical}) and requires further investigations. 

\begin{figure}[h]
\begin{subfigure}{.5\textwidth}
\includegraphics[width=\linewidth]{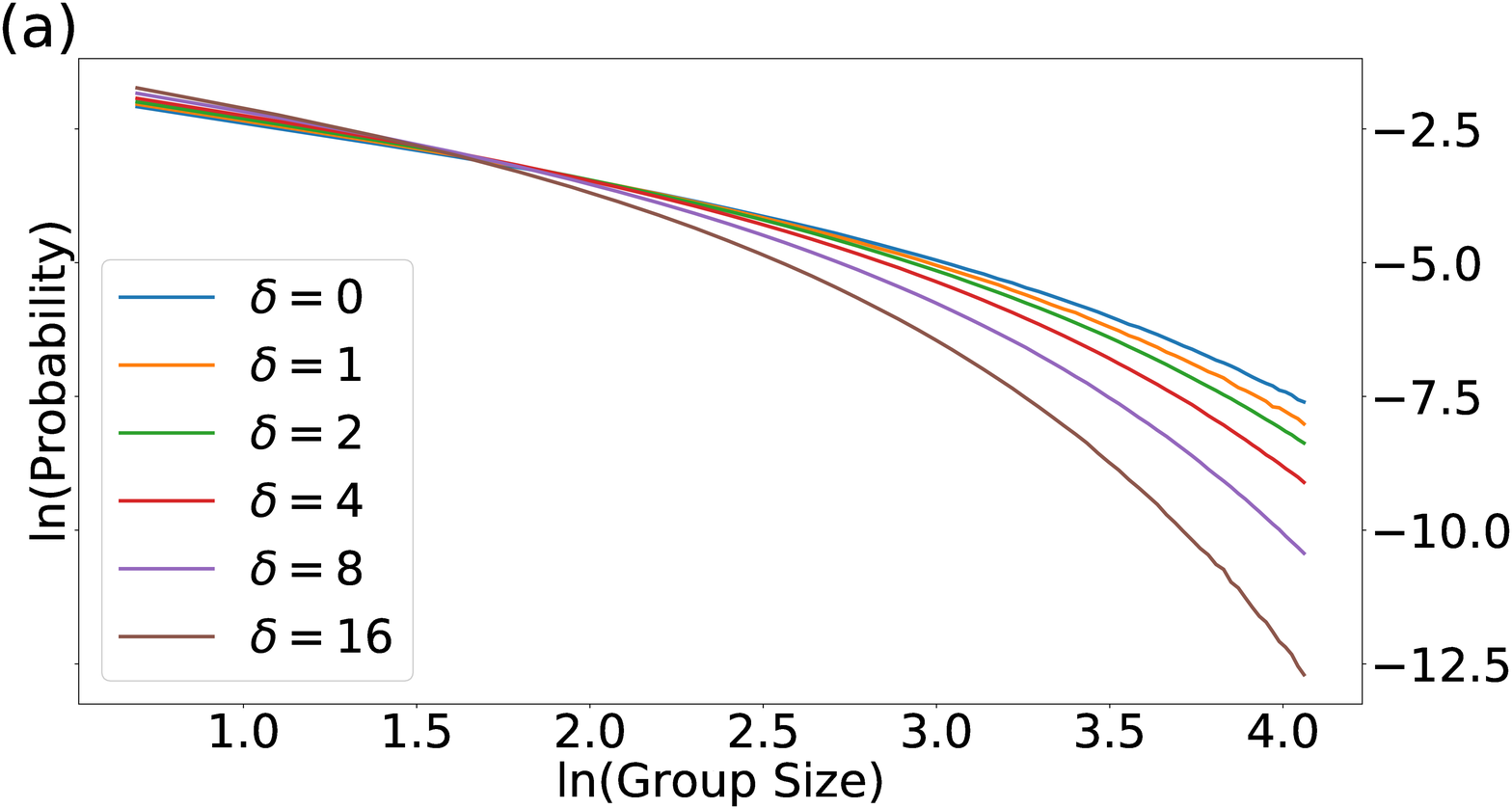}
\end{subfigure}%
\begin{subfigure}{.5\textwidth}
\includegraphics[width=\linewidth]{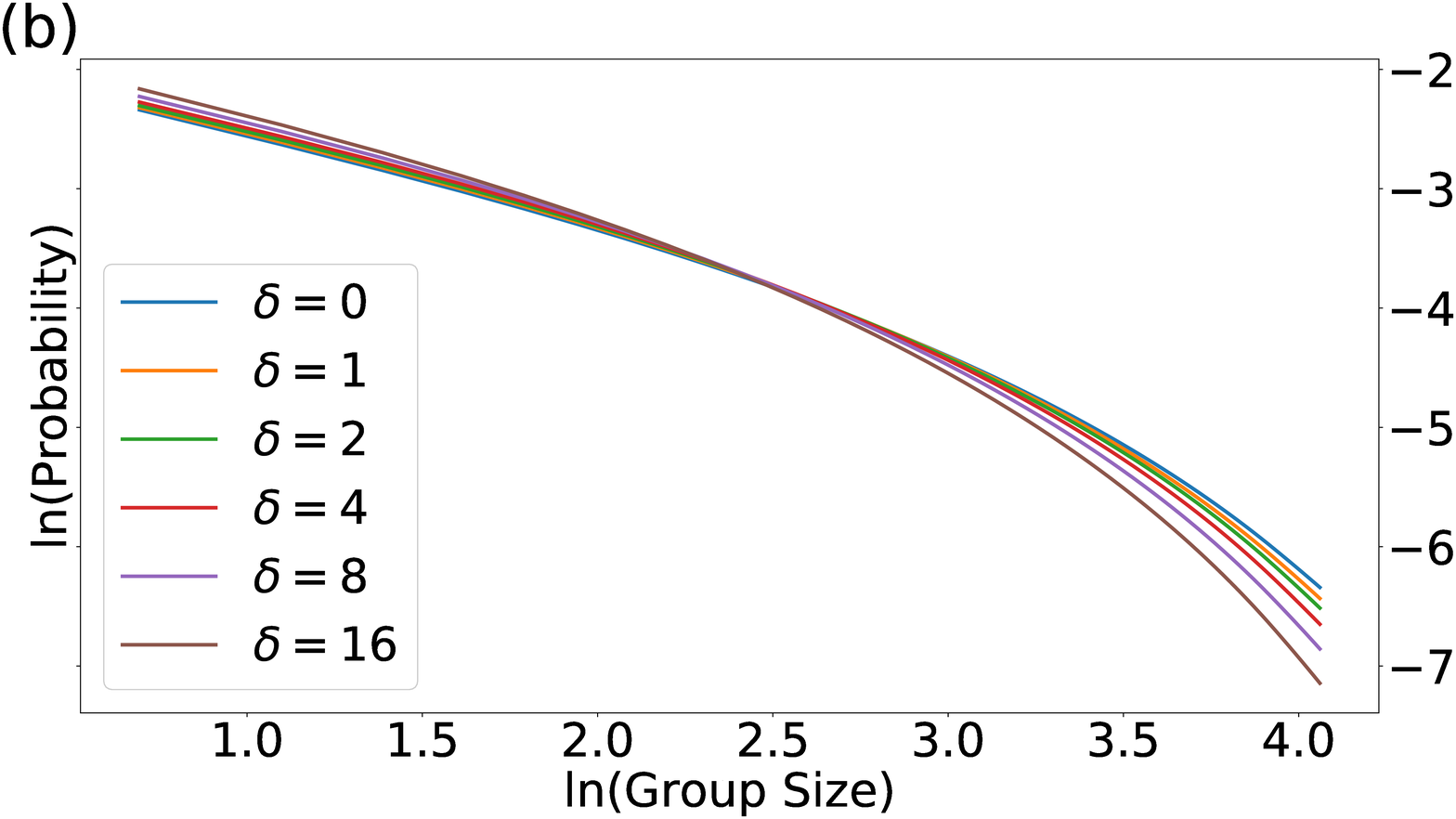}
\end{subfigure}
\caption{Group size distributions plotted in log-log scale. (a) is from Monte-Carlo simulations while (b) is from the analytical model. The two figures show qualitative agreement. The probability as a function of group size decays faster for higher values of $\delta$.}
\label{fig:group_size_distr}
\end{figure}

Earlier studies that adopted merge-split dynamics~\cite{niwa2003JTB,ma2011JTB} primarily investigated the group size distributions in homogeneous populations. We found it instructive to look at the group size distribution for heterogeneous populations too. The probability of a group having size $ n $ is obtained by summing the composition dependent proportion, $ W(n,k) $ over all possible compositions, resulting in a group size probability defined by $ P(n)=\sum_kW(n,k) $. Fig~\ref{fig:group_size_distr} shows $ P(n) $ as a function of $ n $ on log-log scale from simulations and iterative solution to the analytical equations. The plots shows qualitative match with the earlier predicted distributions and is approximately logarithmic. Although the likelihood of occurrence of small groups is nearly the same for different values of excess split-rate for heterogeneous groups ($\delta$), the $P(n)$ decays much faster for larger values of $\delta$. This means that large groups are rarer for higher values of $\delta$. 

\section{Discussion}\label{section:discussions}
In summary, we develop and analyze a heterogeneous flocking model with two types (or species) of individuals.  To the best of our knowledge, this is the first model of merge-split dynamics for heterogeneous populations. We use a first principles approach to derive an analytical description of group sizes and composition. We assumed that heterogeneous groups split at higher rates than homogeneous ones but the rates are independent of the size of the groups. Merge rates are independent of both group size and composition. Our key prediction is that composition of small groups is likely to be skewed towards the abundant type. Above a critical group size, $ n_c $, groups reflect the relative composition of species in the population, i.e.~they are more likely to be heterogeneous. This is despite the assumption that heterogeneous groups split more often. %We discuss the ecological implications of these results below. 

We offer an intuitive explanation of the result via two opposing `forces' at play in this model. The first being chance, driven by the number of combinatorial ways a group can be realised by randomly choosing individuals from the population. Given a heterogeneous population, the combinations for the formation of heterogeneous groups far outweigh that of homogeneous ones; this effect is pronounced when the group size is large. The second force which opposes this formation (or maintenance) of heterogeneous group arises from the model assumption that heterogeneous groups are more likely to split into two daughter groups. A single split, however, is not biased towards formation of homogeneous groups. Nevertheless, successive splits have a cumulative effect of homogenising, and reducing the size of daughter groups.  Therefore, this homogenising force manifests strongly for smaller group sizes. These forces put together, we find the occurrence of homogeneous groups are dominant up to a critical group size $n_c$, beyond which the combinatorial forces result in heterogeneous groups.

In finite groups where individuals probabilistically interact among themselves, the noise at the group-level, also called intrinsic noise, increases with decreasing group size~\cite{van1992stochastic}. Intrinsic noise, in some cases, can cause bimodal states for small groups~\cite{horsthemke1984noise,biancalani2014noise,jhawar2018deriving}. It may be worth investigating a plausible connection between our results, where stochasticity of merge and split events for small groups sizes plays an important role, with the phenomenon of noise-induced bimodality.

\subsection*{Generality and Extensions}

%We see that this assumption holds  true  in the Monte-Carlo simulations of the model even when consider smaller population size ($N=500$) and the number of sites, $s$, is less than the population size ($s=500, 250$). We also find that the qualitative aspects of the results, such as those shown in Figures 3 and 5, continues to hold true for these smaller values of $s$.
 
%However it is important to note that both the assumptions of $s\rightarrow\infty$ and $N\rightarrow\infty$ are necessary for the analysis to work. If either of the assumptions is broken, we may get spurious results. In the case when $s$ is finite, the model is ill-defined since it requires that when a group splits, the daughter groups always have unoccupied sites to move to. On the other hand, if $N$ is kept finite, the probability of merge events becomes so insignificant, that most individuals are solitary.	 

We now discuss some implications of the assumptions of our model and the associated analytical approximations. Our assumption that heterogeneous groups are more likely to split or fragment is broadly supported by previous agent-based simulation models of group movement~\cite{gueron1996jtb,couzin2005nature,del2018PhilTransRoySoc}. However, these models as well as some empirical studies~\cite{ward2008quorum} suggest that, unlike our model assumption, groups do not split uniformly randomly into any of possible partitions; rather, fission events are more likely to cause homogeneous daughter groups. We suspect that incorporating this additional feature, for example in the analysis leading to Fig~\ref{fig:bifurcation_n}, will increase the critical group size ($n_c$), at which the group compositions transition from homogeneous to heterogeneous groups. In other words, we are likely to find groups dominated by the abundant type for much larger groups than predicted by our analyses. Nevertheless, we expect that the qualitative features of our results are unlikely to change.

We made a number of assumptions to derive a semi-analytic approximation for the dynamical equations for the fission-fusion groups. The major ones among these are that the population ($N$) is very large and well mixed and that individuals/groups have a sufficiently large number of sites ($s$) to occupy. Furthermore, we assumed that the proportion of occupied sites ($Z/s$) takes a constant value (i.e. fluctuations in $Z$ are of a smaller order than $s$) in steady state. Another aspect that is implicit in our model formulation is the lack of spatial structure; we assumed, as in the original model by Niwa~\cite{niwa2003JTB,ma2011JTB}, a group in any site can merge with group at any other site and that daughter groups after a split event can occupy any empty site. However, incorporation of such realistic features may make our model analytically intractable.  Therefore, to confirm our predictions in such relatively complicated scenarios, we suggest studies based on individual-based simulations of fission-fusion group dynamics.  

A natural generalisation of our model is one that incorporates $M$ species, with $M>2$. The split rate function for groups could be extrapolated from Eq~\eqref{eq:split_rate} in a way that preserves its qualitative aspects, i.e.~heterogeneous groups having higher split rates than homogeneous ones. For such a system, we expect to find qualitatively similar behavior to that exhibited by the two species model, i.e.~smaller groups are likely to be dominated by one of the species but groups beyond a critical size to be mixed in ratios that are representative of the population composition. To investigate the type of bifurcations and the behavior of the system near critical points, we require a formal analysis of the generalized model. 

\subsection*{Empirical Implications}
We now discuss implications of our results to ecological studies on mixed-species flocks, one of the most widely studied type of heterogeneous flocks. Our model predicts that a study of mixed-species flocks focussing on groups smaller than critical group size of the system will yield observation of flocks that are largely homogeneous; this is despite the fact that the population is heterogeneous. On the other hand, a study on large groups will find flock compositions that represent the population heterogeneity. Therefore, empirical study designs must account for group-size dependent composition of flocks. 

The above prediction of our model has further implications for empirical studies that try to infer interspecies interactions from the frequency of their co-occurrence in mixed-species groups. In such studies, typically, a high frequency of co-occurrence beyond what is expected of a null association is typically interpreted as evidence for positive interspecies interactions~\cite{sridhar2012amnat,berry2014frontier,graves1993pnas,ulrich2010ecology,sridhar2014BehavEcolSocio}. A study that samples flocks that are of size smaller than $n_c$ may rarely find mixed-species associations, thus leading to the conclusion that two species have no or weak positive interspecies interactions. In contrast, a study that samples flocks that are larger than $n_c$ will find many groups with mixed-species associations and thus may arrive at the opposite conclusion of positive interspecies interactions. Therefore, our study highlights that the merge-split dynamics of flocks must be accounted for when making inferences on interspecies interactions.

\subsection*{Conclusions}

Our model analysis yields interesting predictions about the composition of heterogeneous groups, suggesting that groups below a certain threshold do not reflect the population level composition. An interesting direction for further study would be to generalize the model of multiple species to allow for differential interactions between species. The differential interactions may arise because the degree of affinities for different pairs of species are not the same. For example, some species may like to be associated with each other while some may avoid each other. 
Our model provides a starting point to investigate such complex interactions via suitably modified merge and split rate functions. In conclusion, our study highlights the importance of investigating mechanistic models of how individual level interactions between species results in heterogeneous flock dynamics and compositions.  

\section{Acknowledgements}\label{section:acknowledgements}
We thank Hari Sridhar for comments on the manuscript. VG acknowledges support from DBT-IISc partnership program, DST Centre for Mathematical Biology at IISc Phase II (SR/S4/MS:799/12) and infrastructure support from DST-FIST. SK acknowledges partial support from UGC CAS.

\appendix
\section{Fission-fusion dynamics of homogeneous populations}\label{appendix:review}
Our formulation of the problem in heterogeneous populations is based on the model originally conceived by Niwa~\cite{niwa2003JTB} 
%(referred to as Niwa's model henceforth) ~Gokul: we never refer to it as 'Niwa's model' later  
%Vishu: I have changed the title of this section. 
and later analyzed by Ma et al~\cite{ma2011JTB}. To keep the paper self contained as well as to keep it easier to understand the more involved derivations for the heterogeneous case, we review the homogeneous model and its assumptions in this appendix. The model assumes $ s $ sites with no geometry and a population of $ N $ indistinguishable individuals which can occupy these sites. A group is defined to be the set of individuals occupying the same site at any point of time. All groups move at rate $ q $ (which will often be referred to as merge rate) and split at rate $ p $. These rates are independent of group size. When a group moves to an occupied site, they merge to form a larger group with size equal to the sum of smaller groups. A split results in the formation of smaller groups that move to random empty sites. The model can be thought of as the coarse-grained version of a microscopic model with local interactions. 
%Since in the underlying model individuals are oblivious about the size of the group they belong to, it is reasonable to assume that the split and merge rates are independent of group size. 
%Athma: The model description starts abruptly after Ma reference 
%Athma: and a population of N indistinguishable individuals %vishu; accepted.
Ma et al. derived deterministic evolution equations for the merge-split model from first principles. They did so by considering the various changes that could happen to $ f(n,t) $, the expected number of groups of size $ n $ at time $ t $, in a small time interval $ \tau $. Their analysis of the above described merge and split processes resulted in the following coupled differential equations, given below, where $ Z(t):=\sum_{n=1}^{\infty}f(n,t) $ denotes the total number of groups at a given time. 

\begin{equation}
	\begin{aligned}
      	\frac{df(n,t)}{dt} =& \left( \sum_{i=1}^{n-1} qf(i,t)\frac{f(n-i,t)}{s} \right)
        + \sum_{i=n+1}^{\infty} \frac{2pf(i,t)}{i-1}\\ 
      &- 2qf(n,t)\frac{(Z(t)-1)}{s} -\boldsymbol{1}_{n\neq 1}pf(n,t),
      	\end{aligned}
        \tag{A1a}\label{eq:hom_master_eq}
  \end{equation}
  
  \begin{equation}
		\frac{dZ}{dt} = p(Z(t)-f(1,t)) - qZ(t)\frac{(Z(t)-1)}{s},
        \tag{A1b}\label{eq:hom_mean_field}
  \end{equation}

where the symbol $\boldsymbol{1}_{\text{expr}}$ denotes the indicator function, which is 1 when `expr' is satisfied and 0 otherwise. We describe the dynamics that each term of Eq~\eqref{eq:hom_master_eq} and Eq~\eqref{eq:hom_mean_field} represent and a few minor modifications that we propose for better accuracy. The first term captures the event where groups of size $ i $ and $ n-i $ merge, but in the case of $ i = n-i = \frac{n}{2} $, we need to account for over-counting. Furthermore, since a group cannot merge with itself, the term corresponding to $ i =\frac{n}{2} $ has to be $qf(\frac{n}{2},t)\frac{(f(\frac{n}{2},t)-1)}{s}$.

The second term in Eq~\eqref{eq:hom_master_eq} corresponds to larger groups splitting to form groups of size $ n $. The factor of 2 accounts for the fact that a group of size $i>n$ can split in two equally probable ways to yield a group of size $ n $. Since the system is finite, the upper limit in the sum cannot be infinity.

The third term represents the probability of groups with size $ n $ merging with other groups. This can happen in two ways- either the group of size $ n $ moves to an occupied site or a group moves to a site occupied by an $ n $-sized group, hence the factor of two. This term, however, includes merger with groups of size greater than $ N-n $, which is impossible. When we generalize this model to heterogeneous groups, we resolve this issue.	

The fourth term is the decrease in $ f(n,t) $, due to a group of size $ n $ splitting, and does not require modification.

Eq~\eqref{eq:hom_mean_field}, also called the mean-field equation can be obtained by considering the processes that lead to a change in the total number of groups, $Z$. Each split event can increase $Z$ by 1. Since all groups of $n>1$ split at rate $p$, the term is $p(Z(t)-f(1,t))$ is the rate at which $Z$ increases. Each merge event, on the other hand, decreases $Z$ by 1. Since merge events happen when groups move (at rate $q$) to already occupied sites, the total rate associated with merge events is $qZ(t)\frac{(Z(t)-1)}{s}$. The solutions to these equations in steady-state show that fission-fusion dynamics approximately yields a logarithmic group size distribution. We generalize this model and its analytical formulation to heterogeneous populations in the main text.

\section{Deriving the steady state distribution function}\label{appendix:derivation}

%\subsection{Deriving the steady state distribution function}

Here, we present the complete derivation of the steady state equations, Eq~(6) and Eq~(7) from the main text. As in the main text, we denote a group with $n$ individuals of which $k$ are of type-I by the ordered pair $(n,k)$.

\subsection{The Master Equation}\label{appendix:derivation:master}
Groups move from site to site (and effectively merge) at rate $q$. They split at rate $p(n,k)$. The primary random variable of concern is $ X(n,k,t) $, the number of groups with size $ n $, of which $ k $ are of type-I. Let $\mathcal{F}_t = \sigma\{(X(n,k,u):n\geq1, 0 \vee n-N_2 \leq k \leq n \wedge N_1,0\leq u\leq t$\}	denote the $\sigma$-algebra generated by events up to time $ t $ (which contains all the information about the dynamics of the system up to time $t$). Now, we find the expected number of groups of size $n$ with $k$ individuals of type-1 at time $t+\tau$ given information up to time $t$,
\begin{equation}
\begin{aligned}
	&\mathbb{E}\left[X(n,k,t+\tau) | \mathcal{F}_t\right] = (X(n,k,t)+1)Q_{+1}(t)\tau\\ 
    &+ (X(n,k,t)-1)Q_{-1}(t)\tau
    + (X(n,k,t)-2)Q_{-2}(t)\tau
	\nonumber \\
	&+ (X(n,k,t)+1)P_{+1}(t)\tau 
    + (X(n,k,t)-1)P_{-1}(t)\tau\\ 
    &+ (X(n,k,t)+2)P_{+2}(t)\tau +o(\tau) \nonumber
	+ X(n,k,t)(1\\
    &-Q_{+1}(t)\tau
    -Q_{-1}(t)\tau-Q_{-2}(t)\tau
    -P_{-1}(t)\tau -P_{+1}(t)\tau\\
    &-P_{+2}(t)\tau +o(\tau)),
\end{aligned}
\tag{B1}\label{eq:expectation_change}
\end{equation}

where $\lim_{\tau\rightarrow 0}\frac{o(\tau)}{\tau}=0$.

The terms in Eq~\eqref{eq:expectation_change} arise from the following considerations, where we use the notation introduced in the main text, $a\vee b=\max(a,b)$ and $a\wedge b =\min(a,b)$:

	\begin{enumerate}
	
		\item $ Q_{+1}(n,k,t) $: A group, $ (i,j) $ moves to a site occupied by a group $ (n-i,k-j) $ and they merge to produce a larger group $ (n,k) $, thus increasing $X(n,k,t)$ by 1.
%%EDIT in response to minor comment 5
The rate for the event wherein a group $(i,j)$ moves is $qX(i,j,t)$. The probability that a site is occupied by a group $(n-i,k-j)$ is $\frac{X(n-i,k-j,t)}{s}$. Hence the rate for the entire event is given by the product summed over all the possible values of $i$ and $j$. When $n$ and $k$ are even, there is a corner case where two identical groups $(\frac{n}{2},\frac{k}{2})$ merge. Since a group cannot merge with itself the rate will be $qX(\frac{n}{2},\frac{k}{2})\frac{X(\frac{n}{2},\frac{k}{2})-1}{s}$. This yields
        \begin{equation}
		\begin{aligned}
		Q_{+1}(n,k,&t)=\\
        &\Bigg(\sum_{i=1}^{n-1} \sum_{j=(i+k-n) \vee 0}^{i \wedge k} q X(i,j,t) \frac{X(n-i,k-j,t)}{s}\big.\nonumber
		\\&-\Bigg.\boldsymbol{1}_{n\equiv 0(mod~2)} \boldsymbol{1}_{k\equiv 0(mod~2)} q\frac{X(\frac{n}{2},\frac{k}{2},t)}{s} \Bigg).
        \end{aligned}
        \tag{B2}\label{eq:Q_+1}	
        \end{equation}
        
        The second term is accompanied by $\boldsymbol{1}_{n\equiv 0(mod~2)}\boldsymbol{1}_{k\equiv 0(mod~2)}$ (where $a\equiv b(mod~c)$ means that $(a-b)$ is divisible by $c$) imposes the condition that only groups with even $n$ and $k$ can be formed by the merger of identical groups (otherwise $f(n/2,k/2)$ would not make sense, since $n/2$ and $k/2$ would not be integers.
        
        The limits of the sums have been chosen very carefully to account for the finite size of the population. The outer sum goes from $i=1$ to $n-1$, since merge events always result in an increase in group size, groups with size $\geq n$ cannot merge and form a group of size $n$.
        
        The inner sum starts at $j=(i+k-n) \vee 0$. If $(i+k-n)\vee 0 = (i+k-n)$, then $j<(i+k-n)\implies (k-j)>(n-i)$, which would make  $X(n-i,k-j,t)$ meaningless (since a group can't have more type-I individuals than its size). So when $(i+k-n)>0$ we start at $j=(i+k-n)$ instead of $j=0$. 
        
        Similarly, the upper limit is $j=i\wedge k$ because groups with more than $k$ type-I individuals can never produce $(n,k)$ through a merge event. As long as $i<k$ we consider mergers involving groups with up to $i$ type-I individuals, but for $i>k$ we restrict the sum to $k$.
		
		\item $ Q_{-1}(n,k,t) $: A group $ (n,k) $ merges with a group of another size and composition, and decreases the count by one. This term is calculated in a similar way to $Q_{+1}$ except that the group under consideration is merging. It is important to note that the merge can happen in two ways--- the group $(n,k)$ moves to a site occupied by $(i,j)$, and vice-versa. The multiplicative factor of 2 in the first term is to account for this. As reasoned earlier, the second term in the parenthesis accounts for the fact that groups cannot merge with themselves. When a group does merge with an identical group, the change in $X(n,k,t)$ is -2 and not -1, so we need to exclude these kind of events from $Q_{-1}$. The last term that is subtracted accounts for this fact.

\begin{equation}
		\begin{aligned}
		Q_{-1}(n,k,t)=&\\
        &\frac{2q}{s}X(n,k,t) \Bigg( \sum_{i=1}^{N-n}\sum_{j=(n+i-k-N_2) \vee 0}^{(N_1-k) \wedge i}X(i,j,t)\\
        &-\boldsymbol{1}_{n \leq \frac{N}{2}}\boldsymbol{1}_{n-\frac{N_2}{2} \leq k \leq \frac{N1}{2}}X(n,k,t) \Bigg) \nonumber \\
		&-2qX(n,k,t)\frac{X(n,k,t)-1}{s}.
		\end{aligned}
        \tag{B3}\label{eq:Q_-1}
\end{equation}
     
		Groups of size larger than $N/2$ cannot merge with other groups of the same size, since there cannot be more than one group of size greater than $N/2$. Even in groups of size lesser than $N/2$, two groups $(n,k)$ cannot have more than $N_1/2$ or less than $(n-N_2/2)$ type-I individuals. The two indicator functions multiplied with the second term in parenthesis ensures that these constraints are not ignored.
        
        We are considering events where $(n,k)$ groups merge with other groups, therefore groups cannot have sizes more than $N-n$, (the population is finite). The limits of the inner sum, analogously to the one in $Q_{+1}$, restricts the compositions of the groups that can merge. Say $(n+i-k-N_2) > 0$, then $j<(n+i-k-N_2)\implies N_2<(n-k)+(i-j)$, which is impossible, since it would imply that the right hand side of the last inequality, which is the number of type-II individuals in the resulting group is more than total type-II population, $N_2$.
        
		\item $ Q_{-2}(n,k,t) $: Two identical groups $ (n,k) $ merge to give a group $ (2n,2k) $, which results in $ X(n,k,t) $ decreasing by 2. A group $(n,k)$ moves at rate $q$ and lands on a site occupied by an identical group with probability $\frac{X(n,k,t)-1}{s}$, to yield the expression %Also notice that a twice this term is subtracted from \eqref{eq:Q_-1}.
		
		\begin{align}
		Q_{-2}(n,k,t)=qX(n,k,t)\frac{X(n,k,t)-1}{s}.
        \tag{B4}
		\end{align}
		
		\item $ P_{+1}(n,k,t) $: A larger group $ (i,j) $ splits into $ (n,k) $ and $ (i-n,j-k) $ increasing $ X(n,k,t) $ by one. The mechanism of splitting is uniformly at random; consider a group $ (i,j) $; the $ j $ type-I individuals arranged linearly can be split at $ j+1 $ different points and similarly the $ i-j $ type-II individuals can be split in $ i-j+1 $ ways. Combining one part from each of these groups yields the two groups resulting from splitting. However, splits that produce empty groups are not allowed, and this can happen in two extreme cases. Thus the subtraction of 2 from the total is necessary, resulting in $ (j+1)(i-j+1)-2 $ combinations. Since $(i,j)$ splits into $ (n,k) $ and $ (i-n,j-k) $ in two symmetric ways, a multiplicative factor of 2 will arise in the expression.
		
        \begin{equation}
		\begin{aligned}
		P_{+1}(n,k,t) =&\\
        &\sum_{i=n+1}^N \sum_{j=k \vee (i-N_2)}^{(i+k-n) \wedge N_1} \frac{2p(i,j)X(i,j,t)}{(j+1)(i-j+1)-2}\\ 
		&- \frac{2p(2n,2k,t)X(2n,2k,t)}{(2k+1)(2n-2k+1)-2}.
		\end{aligned}
		\tag{B5}\label{eq:P_+1}
        \end{equation}
 
		The second term accounts for the corner case where a group $(2n,2k)$ splits equally to give $(n,k)$-groups. In this case, $X(n,k,t)$ increases by 2 and not 1, and is accounted for in the term below.
        
        In the first term, the index of the outer sum goes from $i=n+1$ to $i=N$, since only groups of size larger than $n$ can split to give groups of size $n$. The inner sum starts at $j=k\vee(i-N_2)$ because when $k>(i-N_2)$, due to the finite number of type-II species, the parent group $(i,j)$ has to have at least $i-N_2$ type-I individuals. The upper limit is $(i+k-n) \wedge N_1$, because $j>(i+k-n)\implies n-k>i-j$, which would imply that the daughter group $(n,k)$ has more type-II individuals than the parent $(i,j)$, which is impossible. 
 			       
		\item $ P_{+2}(n,k,t) $: The event where a group $ (2n,2k) $ splits to give two identical groups $ (n,k) $, increasing $ X(n,k,t) $ by 2. The second term in Eq~\eqref{eq:P_+1} prevents double counting of this event.
		 
		\begin{align}
		P_{+2}(n,k,t) = \frac{p(2n,2k,t)X(2n,2k,t)}{(2k+1)(2n-2k+1)-2}.
        \tag{B6}
        \end{align}
		
		\item $ P_{-1}(n,k,t) $: Group $ (n,k) $ splits. This is the most straightforward of all the events, with a rate given by
		
		\begin{align}
		P_{-1}(n,k,t) = \boldsymbol{1}_{n\neq 1}p(n,k)X(n,k,t).
        \tag{B7}
        \label{eq:P_-1}
		\end{align}
	\end{enumerate}
	
    We can rewrite Eq~\eqref{eq:expectation_change} as follows:

\begin{equation}
    \begin{aligned}
      &\frac{\mathbb{E}\left[X(n,k,t+\tau) | \mathcal{F}_t\right]-X(n,k,t)}{\tau}=Q_{+1}(t)-Q_{-1}(t)\\
      &-2Q_{-2}(t)+P_{+1}(t)-P_{-1}(t)+2P_{+2}(t)+o(1)  
  \end{aligned}
  \tag{B8}\label{eq:expectation_change_rate}
  \end{equation}
  
  Now take the expectation on both sides of Eq~\eqref{eq:expectation_change_rate} and in the continuum limit (i.e.~letting $ \tau\rightarrow 0 $), we get
  
  \begin{equation}
  	\begin{aligned}
    	&\frac{d\mathbb{E}\left[X(n,k,t)\right]}{dt}=\mathbb{E}\left[Q_{+1}(t)\right]-\mathbb{E}\left[Q_{-1}(t)\right]-2\mathbb{E}\left[Q_{-2}(t)\right]\\
        &+\mathbb{E}\left[P_{+1}(t)\right]-\mathbb{E}\left[P_{-1}(t)\right]+2\mathbb{E}\left[P_{+2}(t)\right].
    \end{aligned}
    \tag{B9}\label{eq:het_master_eq_dependent_1}
    \end{equation}
    
\subsection{Steady-State Equation}
    To estimate the expected number of groups at steady state, we set $ \frac{d\mathbb{E}\left[X(n,k,t)\right]}{dt}=0 $, to obtain
  %\begin{widetext}

    \begin{equation}
    \begin{aligned}
    	\mathbb{E}\left[Q_{+1}\right]-\mathbb{E}\left[Q_{-1}\right]-2\mathbb{E}\left[Q_{-2}\right]&\\
        +\mathbb{E}\left[P_{+1}\right]-\mathbb{E}\left[P_{-1}\right]+2\mathbb{E}\left[P_{+2}\right]&=0.
        \end{aligned}
        \tag{B10}\label{eq:ss_expectation}
    \end{equation}
    
  %\end{widetext}
    
    In Eq~\eqref{eq:ss_expectation}, $ P $ and $ Q $, are defined in the same way as before, except that $ X(n,k,t) $ is replaced by $ X(n,k) $, the stationary distribution of the continuous time Markov chain $ \lbrace X(n,k,t)\rbrace $.
    
   To proceed further, we assume that $ N $ is large enough so that the random variables $\{(X(n,k,t):n\geq1, 0 \vee n-N_2 \leq k \leq n \wedge N_1\}$ are pairwise independent. So if we set $ f(n,k):=\mathbb{E}\left[X(n,k)\right] $, then we obtain, from Eq~\eqref{eq:Q_+1}-\eqref{eq:P_-1},Eq~\eqref{eq:ss_expectation},

   \begin{equation}
   \begin{aligned}
   0=&\boldsymbol{1}_{n\neq 1}  \sum_{i=1}^{n-1} \sum_{j=(i+k-n) \vee 0}^{i \wedge k} q f(i,j) \frac{f(n-i,k-j)}{s}\\ 
   &-\boldsymbol{1}_{n\equiv 0(mod~2)} \boldsymbol{1}_{k\equiv 0(mod~2)} q\frac{f(\frac{n}{2},\frac{k}{2})}{s}\nonumber\\ 
   &-\boldsymbol{1}_{n\neq 1}p(n,k)f(n,k)\\
   &+ \sum_{i=n+1}^N \sum_{j=k \vee (i-N_2)}^{(i+k-n) \wedge N_1} \frac{2p(i,j)f(i,j)}{(j+1)(i-j+1)-2}\nonumber\\
   &-\frac{2q}{s}f(n,k)\sum_{i=1}^{N-n}\sum_{j=(n+i-k-N_2) \vee 0}^{(N_1-k) \wedge i}f(i,j)\\
   &+\boldsymbol{1}_{n-\frac{N_2}{2} \leq k \leq \frac{N1}{2}}2q\frac{f(n,k)}{s}.
   \end{aligned}
   \tag{B11}\label{eq:ss_master_equation_indep}
   \end{equation}
   
\subsection{Mean-Field Equation}
\label{appendix:derivation:meanfield}
   We can also derive an equation for the total number of groups, $Z(t)=\sum_n\sum_kf(n,k,t)$ by generalising Eq~\eqref{eq:hom_mean_field}. Each split event increases $Z(t)$ by 1 and each merge event decreases $Z(t)$ by 1. 
   
  The rate of increase of $Z(t)$ due to all the current groups $(i,j)$ is $p(i,j)f(i,j,t)$, hence the total contribution of split events to $\frac{dZ}{dt}$ is this term summed over all valid $i$'s and $j$'s.
  
  At a given time, groups $(i,j)$ merge with $(k,l)$ at a rate $\frac{q}{s}f(i,j,t)f(k,l,t)$, so this term summed over all possible $i$'s, $j$'s, $k$'s, and $l$'s gives the total rate of decrease of $Z(t)$.
  
\begin{equation}   
   \begin{aligned}
 	\frac{dZ(t)}{dt} =& \sum_{i=2}^{N}\sum_{0\vee (i-N_2)}^{i\wedge N_1}p(i,j)f(i,j,t)\\
    &-\sum_{i=1}^{N}\sum_{0\vee(i-N_2)}^{i\wedge N_1}\Bigg( \frac{q}{s}f(i,j,t)\sum_{k=1}^{N-i}\sum_{l=(i+k-j-N_2)\vee 0}^{(N_1-j)\wedge k}\\
    &f(k,l,t)-\boldsymbol{1}_{i\leq \frac{N}{2}}\boldsymbol{1}_{i-\frac{N_2}{2}\leq j\leq \frac{N_1}{2}}\frac{q}{s}f(i,j,t)\Bigg).
 \end{aligned}
 \tag{B12}\label{eq:het_mean_field_1}
\end{equation}

 An additional term is subtracted in the second sum to account for the fact that groups cannot merge with themselves.

\subsection{Scaling Limit}
\label{appendix:derivation:scaling}
   Dividing Eq~\eqref{eq:ss_master_equation_indep} by $ Z^2 $ we can rewrite the equation in terms of $ W(n,k) \equiv \frac{f(n,k)}{Z} $ (the proportion of $(n,k)$-groups), we get
   \begin{equation}
   \begin{aligned}
   0=&\Bigg(\sum_{i=1}^{n-1} \sum_{j=(i+k-n) \vee 0}^{i \wedge k} q W(i,j) \frac{W(n-i,k-j)}{s}\\
   &-\boldsymbol{1}_{n\equiv 0(mod~2)} \boldsymbol{1}_{k\equiv 0(mod~2)} q\frac{W(\frac{n}{2},\frac{k}{2})}{Zs}\Bigg)\nonumber\\ 
   &-\boldsymbol{1}_{n\neq 1}\frac{p(n,k)}{Z}W(n,k)+\boldsymbol{1}_{n-\frac{N_2}{2} \leq k \leq \frac{N1}{2}}2q\frac{W(n,k)}{Zs}\\
   &+ \sum_{i=n+1}^N \sum_{j=k \vee (i-N_2)}^{(i+k-n) \wedge N_1} \frac{2p(i,j)W(i,j)}{Z((j+1)(i-j+1)-2)}\nonumber\\
   &-\frac{2q}{s}W(n,k)\sum_{i=1}^{N-n}\sum_{j=(n+i-k-N_2) \vee 0}^{(N_1-k) \wedge i}W(i,j).
   \end{aligned}
	\tag{B13}\label{eq:ss_master_no_approx}
   \end{equation}
   
 We also divide the steady-state ($\frac{dZ(t)}{dt}=0$) version of Eq~\eqref{eq:het_mean_field_1} by $Z^2$,
 \begin{equation}
 \begin{aligned}
 	0 =& \sum_{i=2}^{N}\sum_{0\vee (i-N_2)}^{i\wedge N_1}\frac{p(i,j)}{Z}W(i,j)\\
    &-\sum_{i=1}^{N}\sum_{0\vee(i-N_2)}^{i\wedge N_1}\Bigg( \frac{q}{s}W(i,j)\sum_{k=1}^{N-i}\sum_{l=(i+k-j-N_2)\vee 0}^{(N_1-j)\wedge k}W(k,l)\nonumber\\
    &+\boldsymbol{1}_{i\leq \frac{N}{2}}\boldsymbol{1}_{i-\frac{N_2}{2}\leq j\leq \frac{N_1}{2}}\frac{q}{Zs}W(i,j)\Bigg).
 \end{aligned}
 \tag{B14}\label{eq:ss_mean_field_no_approx}
\end{equation}

 Since $ N $ is large and $ s\rightarrow \infty $, we must speed up the split and merge rates to obtain a non-trivial limit and consequently set the split and merge rates to $q=qs$ and $p=ps$ in Eq~\eqref{eq:ss_master_no_approx} and Eq~\eqref{eq:ss_mean_field_no_approx}. In this limit, we can use the a priori knowledge that the mass of the group distribution is concentrated at small $n$, to remove the constraints on the last two terms of Eq~\eqref{eq:ss_master_no_approx}. Consequently, the double sum yields $2qW(n,k)$ (since $W(i,j)$ is normalized). We drop the constraint in the last term as well since as we will show below this term will be negligible in the limit and can be ignored.
% It also becomes acceptable to relax the finite size constraint on merge events and replace the combined term $Q_1+2Q_2$ with $qX(n,k,t)(Z(t)-1)$. 

Using these approximations in Eq~\eqref{eq:ss_master_no_approx} gives
\begin{equation}
 \begin{aligned}
 	0=&\boldsymbol{1}_{n\neq 1}  \Bigg(\sum_{i=1}^{n-1} \sum_{j=(i+k-n) \vee 0}^{i \wedge k} q W(i,j) W(n-i,k-j)\\
    &-\boldsymbol{1}_{n\equiv 0(mod~2)} \boldsymbol{1}_{k\equiv 0(mod~2)} q\frac{W(\frac{n}{2},\frac{k}{2})}{Z}\Bigg)\nonumber\\
    &+ \sum_{i=n+1}^N \sum_{j=k \vee (i-N_2)}^{(i+k-n) \wedge N_1} \frac{2p(i,j)W(i,j)}{\frac{Z}{s}((j+1)(i-j+1)-2)}\\
    &-2qW(n,k)(1-\frac{1}{Z})-\boldsymbol{1}_{n\neq 1}\frac{p(n,k)}{\frac{Z}{s}}W(n,k).
 \end{aligned}
 \tag{B15}
\end{equation}

We assume that when the system scales, $Z$ also scales such that $ \frac{Z}{s}\rightarrow Z_0$, (the fraction of occupied sites in steady-state) and consequently we can write the steady state equation for $ W(n,k) $ as
\begin{equation}
 \begin{aligned}
 0=&\sum_{i=1}^{n-1} \sum_{j=(i+k-n) \vee 0}^{i \wedge k} q W(i,j) W(n-i,k-j)\\
 &-\boldsymbol{1}_{n\neq 1}\frac{p(n,k)}{Z_0}W(n,k)-2qW(n,k)\nonumber\\&+\sum_{i=n+1}^N \sum_{j=k \vee (i-N_2)}^{(i+k-n) \wedge N_1} \frac{2p(i,j)W(i,j)}{Z_0((j+1)(i-j+1)-2)}.
 \end{aligned}
 \tag{B16}
 \end{equation}
 
We also incorporate the scaling assumption for $q$ and $p$ into the steady-state mean-field equation, Eq~\eqref{eq:ss_mean_field_no_approx} which gives
\begin{equation}
\begin{aligned}
 	0 =& \sum_{i=2}^{N}\sum_{0\vee (i-N_2)}^{i\wedge N_1}\frac{p(i,j)}{\frac{Z}{s}}W(i,j)\\
    &-q\sum_{i=1}^{N}\sum_{0\vee(i-N_2)}^{i\wedge N_1}\Bigg( W(i,j)\sum_{k=1}^{N-i}\sum_{l=(i+k-j-N_2)\vee 0}^{(N_1-j)\wedge k}W(k,l)\nonumber\\
    &+\boldsymbol{1}_{i\leq \frac{N}{2}}\boldsymbol{1}_{i-\frac{N_2}{2}\leq j\leq \frac{N_1}{2}}\frac{q}{Z}W(i,j)\Bigg).
    \label{eq:ss_master_eq}
 \end{aligned}
 \tag{B17}
\end{equation}

In the large $s$ and $Z$ limit, the second term in parentheses goes to 0. Relaxing the finite size constraint on the first term in parentheses will simplify the equation considerably, since $\sum_n\sum_kW(n,k)=1$, leaving only $q$. Again assuming that $ \frac{Z}{s}\rightarrow Z_0$, we can write

 \begin{equation}
 Z_0 = \frac{1}{q}\sum_{i=2}^{N}\sum_{j=0\vee(i-N_2)}^{i\wedge N_1}p(i,j)W(i,j,t).
 \tag{B18}\label{eq:ss_mean_field_eq}
 \end{equation}	
 
\textbf{Remarks:} It is important to note that, although in the absence of excess split-rate for heterogeneous groups (i.e. $ \delta=0$), the split rate reduces to the homogeneous case, $ g(n,t):= \sum_k\mathbb{E}\left[X(n,k,t)\right] $ will not obey the homogeneous master equation Eq~\eqref{eq:hom_master_eq}. This is because in Eq~(2), from the main text, the terms $ P_{+1} $ and $ P_{+2} $ are derived by assuming that individuals of the same type are indistinguishable, but of different types are distinguishable. However, in the homogeneous case, all individuals are indistinguishable. Hence, if both the equations were derived assuming that all individuals are distinguishable, $ g(n,t) $ will obey the homogeneous equation also.

\subsection{Iterative Solution}
\label{appendix:derivation:iterative}
It is a challenging task to solve Eq~\eqref{eq:ss_master_eq} and Eq~\eqref{eq:ss_mean_field_eq} directly, so we use an iterative scheme to obtain the solutions. Eq~\eqref{eq:ss_master_eq} can be used to express the proportion of groups of a particular size and composition in terms of the proportions of groups of other sizes and compositions.

\begin{equation}
\begin{aligned}
&W^{(m+1)}(n,k)=\\
&\Bigg(\frac{p(n,k)}{Z_0^{(m)}}+2q\Bigg)^{-1}\\
&\Bigg(\boldsymbol{1}_{n\neq1}\sum_{i=1}^{n-1} \sum_{j=(i+k-n) \vee 0}^{i \wedge k} q W^{(m)}(i,j) W^{(m)}(n-i,k-j)\nonumber\\&+\sum_{i=n+1}^N \sum_{j=k \vee (i-N_2)}^{(i+k-n) \wedge N_1} \frac{2p(i,j)W^{(m)}(i,j)}{Z_0^{(m)}((j+1)(i-j+1)-2)}\Bigg),
\end{aligned}
\tag{B19a}
\end{equation}

\begin{equation}
\begin{aligned}
Z_0^{(m)} := \frac{1}{q}\sum_{i=2}^{N}\sum_{j=0\vee(i-N_2)}^{i\wedge N_1}p(i,j)W^{(m)}(i,j,t). 
\end{aligned}
\tag{B19b}
\end{equation}

$m$ is the index of iteration. As the initial condition for the fixed point iteration we use the solution to the homogeneous equation ($g(n)$). We distribute $g(n)$ uniformly to $f^1(n,k)$, for $ N_1=N_2 $. This set of iterative equations goes to steady state, and produces stable solutions. %However, as a result of the approximations we adopted, the total population size is not conserved during the iteration.

%The probability that an individual picked at random belongs to a group of size $ n $ is proportional to $ nP(n) $ and can be interpreted as grouping preference of individuals. We normalized and plotted $ nP(n) $ against $ n $. Although for large $ n $, there is a qualitative match between the results of the simulation and the analytical approximation, the similarity breaks down for small values of $ n $, as is evident from the fact the latter shows a maxima while the former does not.

%\onecolumngrid

\bibliographystyle{apsrev4-1}
\bibliography{master_bibliography}

\begin{thebibliography}{55}%
\makeatletter
\providecommand \@ifxundefined [1]{%
 \@ifx{#1\undefined}
}%
\providecommand \@ifnum [1]{%
 \ifnum #1\expandafter \@firstoftwo
 \else \expandafter \@secondoftwo
 \fi
}%
\providecommand \@ifx [1]{%
 \ifx #1\expandafter \@firstoftwo
 \else \expandafter \@secondoftwo
 \fi
}%
\providecommand \natexlab [1]{#1}%
\providecommand \enquote  [1]{``#1''}%
\providecommand \bibnamefont  [1]{#1}%
\providecommand \bibfnamefont [1]{#1}%
\providecommand \citenamefont [1]{#1}%
\providecommand \href@noop [0]{\@secondoftwo}%
\providecommand \href [0]{\begingroup \@sanitize@url \@href}%
\providecommand \@href[1]{\@@startlink{#1}\@@href}%
\providecommand \@@href[1]{\endgroup#1\@@endlink}%
\providecommand \@sanitize@url [0]{\catcode `\\12\catcode `\$12\catcode
  `\&12\catcode `\#12\catcode `\^12\catcode `\_12\catcode `\%12\relax}%
\providecommand \@@startlink[1]{}%
\providecommand \@@endlink[0]{}%
\providecommand \url  [0]{\begingroup\@sanitize@url \@url }%
\providecommand \@url [1]{\endgroup\@href {#1}{\urlprefix }}%
\providecommand \urlprefix  [0]{URL }%
\providecommand \Eprint [0]{\href }%
\providecommand \doibase [0]{http://dx.doi.org/}%
\providecommand \selectlanguage [0]{\@gobble}%
\providecommand \bibinfo  [0]{\@secondoftwo}%
\providecommand \bibfield  [0]{\@secondoftwo}%
\providecommand \translation [1]{[#1]}%
\providecommand \BibitemOpen [0]{}%
\providecommand \bibitemStop [0]{}%
\providecommand \bibitemNoStop [0]{.\EOS\space}%
\providecommand \EOS [0]{\spacefactor3000\relax}%
\providecommand \BibitemShut  [1]{\csname bibitem#1\endcsname}%
\let\auto@bib@innerbib\@empty
%</preamble>
\bibitem [{\citenamefont {Parrish}\ and\ \citenamefont
  {Edelstein-Keshet}(1999)}]{parrish1999science}%
  \BibitemOpen
  \bibfield  {author} {\bibinfo {author} {\bibfnamefont {J.~K.}\ \bibnamefont
  {Parrish}}\ and\ \bibinfo {author} {\bibfnamefont {L.}~\bibnamefont
  {Edelstein-Keshet}},\ }\href@noop {} {\bibfield  {journal} {\bibinfo
  {journal} {Science}\ }\textbf {\bibinfo {volume} {284}},\ \bibinfo {pages}
  {99} (\bibinfo {year} {1999})}\BibitemShut {NoStop}%
\bibitem [{\citenamefont {Camazine}(2003)}]{camazine2003Book}%
  \BibitemOpen
  \bibfield  {author} {\bibinfo {author} {\bibfnamefont {S.}~\bibnamefont
  {Camazine}},\ }\href@noop {} {\emph {\bibinfo {title} {Self-organization in
  biological systems}}}\ (\bibinfo  {publisher} {Princeton University Press},\
  \bibinfo {year} {2003})\BibitemShut {NoStop}%
\bibitem [{\citenamefont {Sumpter}(2010)}]{sumpter2010Book}%
  \BibitemOpen
  \bibfield  {author} {\bibinfo {author} {\bibfnamefont {D.~J.}\ \bibnamefont
  {Sumpter}},\ }\href@noop {} {\emph {\bibinfo {title} {Collective animal
  behavior}}}\ (\bibinfo  {publisher} {Princeton University Press},\ \bibinfo
  {year} {2010})\BibitemShut {NoStop}%
\bibitem [{\citenamefont {Couzin}\ and\ \citenamefont
  {Krause}(2003)}]{couzin2003adv}%
  \BibitemOpen
  \bibfield  {author} {\bibinfo {author} {\bibfnamefont {I.~D.}\ \bibnamefont
  {Couzin}}\ and\ \bibinfo {author} {\bibfnamefont {J.}~\bibnamefont
  {Krause}},\ }\href@noop {} {\bibfield  {journal} {\bibinfo  {journal}
  {Advances in the Study of Behavior}\ }\textbf {\bibinfo {volume} {32}},\
  \bibinfo {pages} {1} (\bibinfo {year} {2003})}\BibitemShut {NoStop}%
\bibitem [{\citenamefont {Vicsek}\ \emph {et~al.}(1995)\citenamefont {Vicsek},
  \citenamefont {Czir{\'o}k}, \citenamefont {Ben-Jacob}, \citenamefont
  {Cohen},\ and\ \citenamefont {Shochet}}]{vicsek1995prl}%
  \BibitemOpen
  \bibfield  {author} {\bibinfo {author} {\bibfnamefont {T.}~\bibnamefont
  {Vicsek}}, \bibinfo {author} {\bibfnamefont {A.}~\bibnamefont {Czir{\'o}k}},
  \bibinfo {author} {\bibfnamefont {E.}~\bibnamefont {Ben-Jacob}}, \bibinfo
  {author} {\bibfnamefont {I.}~\bibnamefont {Cohen}}, \ and\ \bibinfo {author}
  {\bibfnamefont {O.}~\bibnamefont {Shochet}},\ }\href@noop {} {\bibfield
  {journal} {\bibinfo  {journal} {Physical review letters}\ }\textbf {\bibinfo
  {volume} {75}},\ \bibinfo {pages} {1226} (\bibinfo {year}
  {1995})}\BibitemShut {NoStop}%
\bibitem [{\citenamefont {del Mar~Delgado}\ \emph {et~al.}(2018)\citenamefont
  {del Mar~Delgado}, \citenamefont {Miranda}, \citenamefont {Alvarez},
  \citenamefont {Gurarie}, \citenamefont {Fagan}, \citenamefont {Penteriani},
  \citenamefont {di~Virgilio},\ and\ \citenamefont
  {Morales}}]{del2018PhilTransRoySoc}%
  \BibitemOpen
  \bibfield  {author} {\bibinfo {author} {\bibfnamefont {M.}~\bibnamefont {del
  Mar~Delgado}}, \bibinfo {author} {\bibfnamefont {M.}~\bibnamefont {Miranda}},
  \bibinfo {author} {\bibfnamefont {S.~J.}\ \bibnamefont {Alvarez}}, \bibinfo
  {author} {\bibfnamefont {E.}~\bibnamefont {Gurarie}}, \bibinfo {author}
  {\bibfnamefont {W.~F.}\ \bibnamefont {Fagan}}, \bibinfo {author}
  {\bibfnamefont {V.}~\bibnamefont {Penteriani}}, \bibinfo {author}
  {\bibfnamefont {A.}~\bibnamefont {di~Virgilio}}, \ and\ \bibinfo {author}
  {\bibfnamefont {J.~M.}\ \bibnamefont {Morales}},\ }\href@noop {} {\bibfield
  {journal} {\bibinfo  {journal} {Phil. Trans. R. Soc. B}\ }\textbf {\bibinfo
  {volume} {373}},\ \bibinfo {pages} {20170008} (\bibinfo {year}
  {2018})}\BibitemShut {NoStop}%
\bibitem [{\citenamefont {Dyer}\ \emph {et~al.}(2008)\citenamefont {Dyer},
  \citenamefont {Croft}, \citenamefont {Morrell},\ and\ \citenamefont
  {Krause}}]{dyer2008BehavEcol}%
  \BibitemOpen
  \bibfield  {author} {\bibinfo {author} {\bibfnamefont {J.~R.}\ \bibnamefont
  {Dyer}}, \bibinfo {author} {\bibfnamefont {D.~P.}\ \bibnamefont {Croft}},
  \bibinfo {author} {\bibfnamefont {L.~J.}\ \bibnamefont {Morrell}}, \ and\
  \bibinfo {author} {\bibfnamefont {J.}~\bibnamefont {Krause}},\ }\href@noop {}
  {\bibfield  {journal} {\bibinfo  {journal} {Behavioral Ecology}\ }\textbf
  {\bibinfo {volume} {20}},\ \bibinfo {pages} {165} (\bibinfo {year}
  {2008})}\BibitemShut {NoStop}%
\bibitem [{\citenamefont {Ioannou}\ \emph {et~al.}(2008)\citenamefont
  {Ioannou}, \citenamefont {Payne},\ and\ \citenamefont
  {Krause}}]{ioannou2008Oecol}%
  \BibitemOpen
  \bibfield  {author} {\bibinfo {author} {\bibfnamefont {C.}~\bibnamefont
  {Ioannou}}, \bibinfo {author} {\bibfnamefont {M.}~\bibnamefont {Payne}}, \
  and\ \bibinfo {author} {\bibfnamefont {J.}~\bibnamefont {Krause}},\
  }\href@noop {} {\bibfield  {journal} {\bibinfo  {journal} {Oecologia}\
  }\textbf {\bibinfo {volume} {157}},\ \bibinfo {pages} {177} (\bibinfo {year}
  {2008})}\BibitemShut {NoStop}%
\bibitem [{\citenamefont {Michelena}\ \emph {et~al.}(2010)\citenamefont
  {Michelena}, \citenamefont {Jeanson}, \citenamefont {Deneubourg},\ and\
  \citenamefont {Sibbald}}]{michelena2010ProcRoySocB}%
  \BibitemOpen
  \bibfield  {author} {\bibinfo {author} {\bibfnamefont {P.}~\bibnamefont
  {Michelena}}, \bibinfo {author} {\bibfnamefont {R.}~\bibnamefont {Jeanson}},
  \bibinfo {author} {\bibfnamefont {J.-L.}\ \bibnamefont {Deneubourg}}, \ and\
  \bibinfo {author} {\bibfnamefont {A.~M.}\ \bibnamefont {Sibbald}},\
  }\href@noop {} {\bibfield  {journal} {\bibinfo  {journal} {Proceedings of the
  Royal Society of London B: Biological Sciences}\ }\textbf {\bibinfo {volume}
  {277}},\ \bibinfo {pages} {1093} (\bibinfo {year} {2010})}\BibitemShut
  {NoStop}%
\bibitem [{\citenamefont {Morse}(1970)}]{morse1970EcolMono}%
  \BibitemOpen
  \bibfield  {author} {\bibinfo {author} {\bibfnamefont {D.~H.}\ \bibnamefont
  {Morse}},\ }\href@noop {} {\bibfield  {journal} {\bibinfo  {journal}
  {Ecological monographs}\ }\textbf {\bibinfo {volume} {40}},\ \bibinfo {pages}
  {119} (\bibinfo {year} {1970})}\BibitemShut {NoStop}%
\bibitem [{\citenamefont {Diamond}(1981)}]{diamond1981Nature}%
  \BibitemOpen
  \bibfield  {author} {\bibinfo {author} {\bibfnamefont {J.~M.}\ \bibnamefont
  {Diamond}},\ }\href@noop {} {\bibfield  {journal} {\bibinfo  {journal}
  {Nature}\ }\textbf {\bibinfo {volume} {292}},\ \bibinfo {pages} {408}
  (\bibinfo {year} {1981})}\BibitemShut {NoStop}%
\bibitem [{\citenamefont {Sridhar}\ \emph {et~al.}(2009)\citenamefont
  {Sridhar}, \citenamefont {Beauchamp},\ and\ \citenamefont
  {Shanker}}]{sridhar2009animalbeh}%
  \BibitemOpen
  \bibfield  {author} {\bibinfo {author} {\bibfnamefont {H.}~\bibnamefont
  {Sridhar}}, \bibinfo {author} {\bibfnamefont {G.}~\bibnamefont {Beauchamp}},
  \ and\ \bibinfo {author} {\bibfnamefont {K.}~\bibnamefont {Shanker}},\
  }\href@noop {} {\bibfield  {journal} {\bibinfo  {journal} {Animal Behaviour}\
  }\textbf {\bibinfo {volume} {78}},\ \bibinfo {pages} {337} (\bibinfo {year}
  {2009})}\BibitemShut {NoStop}%
\bibitem [{\citenamefont {Stensland}\ \emph {et~al.}(2003)\citenamefont
  {Stensland}, \citenamefont {Angerbj{\"o}rn},\ and\ \citenamefont
  {Berggren}}]{stensland2003mammalreview}%
  \BibitemOpen
  \bibfield  {author} {\bibinfo {author} {\bibfnamefont {E.}~\bibnamefont
  {Stensland}}, \bibinfo {author} {\bibfnamefont {A.}~\bibnamefont
  {Angerbj{\"o}rn}}, \ and\ \bibinfo {author} {\bibfnamefont {P.}~\bibnamefont
  {Berggren}},\ }\href@noop {} {\bibfield  {journal} {\bibinfo  {journal}
  {Mammal Review}\ }\textbf {\bibinfo {volume} {33}},\ \bibinfo {pages} {205}
  (\bibinfo {year} {2003})}\BibitemShut {NoStop}%
\bibitem [{\citenamefont {Greenberg}(2000)}]{greenberg2000book}%
  \BibitemOpen
  \bibfield  {author} {\bibinfo {author} {\bibfnamefont {R.}~\bibnamefont
  {Greenberg}},\ }\href@noop {} {\bibfield  {journal} {\bibinfo  {journal} {On
  the move: How and why animals travel in groups}\ ,\ \bibinfo {pages} {521}}
  (\bibinfo {year} {2000})}\BibitemShut {NoStop}%
\bibitem [{\citenamefont {Lukoschek}\ and\ \citenamefont
  {McCormick}(2000)}]{lukoschek2000review}%
  \BibitemOpen
  \bibfield  {author} {\bibinfo {author} {\bibfnamefont {V.}~\bibnamefont
  {Lukoschek}}\ and\ \bibinfo {author} {\bibfnamefont {M.}~\bibnamefont
  {McCormick}},\ }in\ \href@noop {} {\emph {\bibinfo {booktitle} {Proceeding
  9th International Coral Reef Symposium}}},\ Vol.~\bibinfo {volume} {1}\
  (\bibinfo {organization} {Ministry of Environment, Indonesian Institute of
  Sciences and International Society for Reef Studies},\ \bibinfo {year}
  {2000})\ pp.\ \bibinfo {pages} {467--474}\BibitemShut {NoStop}%
\bibitem [{\citenamefont {Sridhar}\ and\ \citenamefont
  {Guttal}(2018)}]{sridhar2018philB}%
  \BibitemOpen
  \bibfield  {author} {\bibinfo {author} {\bibfnamefont {H.}~\bibnamefont
  {Sridhar}}\ and\ \bibinfo {author} {\bibfnamefont {V.}~\bibnamefont
  {Guttal}},\ }\href@noop {} {\bibfield  {journal} {\bibinfo  {journal}
  {Philosophical Transactions of the Royal Society B: Biological Sciences}\
  }\textbf {\bibinfo {volume} {373}},\ \bibinfo {pages} {20170014} (\bibinfo
  {year} {2018})}\BibitemShut {NoStop}%
\bibitem [{\citenamefont {Gueron}\ \emph {et~al.}(1996)\citenamefont {Gueron},
  \citenamefont {Levin},\ and\ \citenamefont {Rubenstein}}]{gueron1996jtb}%
  \BibitemOpen
  \bibfield  {author} {\bibinfo {author} {\bibfnamefont {S.}~\bibnamefont
  {Gueron}}, \bibinfo {author} {\bibfnamefont {S.~A.}\ \bibnamefont {Levin}}, \
  and\ \bibinfo {author} {\bibfnamefont {D.~I.}\ \bibnamefont {Rubenstein}},\
  }\href@noop {} {\bibfield  {journal} {\bibinfo  {journal} {Journal of
  Theoretical Biology}\ }\textbf {\bibinfo {volume} {182}},\ \bibinfo {pages}
  {85} (\bibinfo {year} {1996})}\BibitemShut {NoStop}%
\bibitem [{\citenamefont {Nagy}\ \emph {et~al.}(2010)\citenamefont {Nagy},
  \citenamefont {Akos}, \citenamefont {Biro},\ and\ \citenamefont
  {Vicsek}}]{nagy2010nature}%
  \BibitemOpen
  \bibfield  {author} {\bibinfo {author} {\bibfnamefont {M.}~\bibnamefont
  {Nagy}}, \bibinfo {author} {\bibfnamefont {Z.}~\bibnamefont {Akos}}, \bibinfo
  {author} {\bibfnamefont {D.}~\bibnamefont {Biro}}, \ and\ \bibinfo {author}
  {\bibfnamefont {T.}~\bibnamefont {Vicsek}},\ }\href@noop {} {\bibfield
  {journal} {\bibinfo  {journal} {Nature}\ }\textbf {\bibinfo {volume} {464}},\
  \bibinfo {pages} {890} (\bibinfo {year} {2010})}\BibitemShut {NoStop}%
\bibitem [{\citenamefont {Biro}\ \emph {et~al.}(2006)\citenamefont {Biro},
  \citenamefont {Sumpter}, \citenamefont {Meade},\ and\ \citenamefont
  {Guilford}}]{biro2006currbio}%
  \BibitemOpen
  \bibfield  {author} {\bibinfo {author} {\bibfnamefont {D.}~\bibnamefont
  {Biro}}, \bibinfo {author} {\bibfnamefont {D.~J.}\ \bibnamefont {Sumpter}},
  \bibinfo {author} {\bibfnamefont {J.}~\bibnamefont {Meade}}, \ and\ \bibinfo
  {author} {\bibfnamefont {T.}~\bibnamefont {Guilford}},\ }\href@noop {}
  {\bibfield  {journal} {\bibinfo  {journal} {Current Biology}\ }\textbf
  {\bibinfo {volume} {16}},\ \bibinfo {pages} {2123} (\bibinfo {year}
  {2006})}\BibitemShut {NoStop}%
\bibitem [{\citenamefont {Romey}\ and\ \citenamefont
  {Vidal}(2013)}]{romey2013ecolmodel}%
  \BibitemOpen
  \bibfield  {author} {\bibinfo {author} {\bibfnamefont {W.~L.}\ \bibnamefont
  {Romey}}\ and\ \bibinfo {author} {\bibfnamefont {J.~M.}\ \bibnamefont
  {Vidal}},\ }\href@noop {} {\bibfield  {journal} {\bibinfo  {journal}
  {Ecological modelling}\ }\textbf {\bibinfo {volume} {258}},\ \bibinfo {pages}
  {9} (\bibinfo {year} {2013})}\BibitemShut {NoStop}%
\bibitem [{\citenamefont {Herbert-Read}\ \emph {et~al.}(2013)\citenamefont
  {Herbert-Read}, \citenamefont {Krause}, \citenamefont {Morrell},
  \citenamefont {Schaerf}, \citenamefont {Krause},\ and\ \citenamefont
  {Ward}}]{herbert2013procB}%
  \BibitemOpen
  \bibfield  {author} {\bibinfo {author} {\bibfnamefont {J.~E.}\ \bibnamefont
  {Herbert-Read}}, \bibinfo {author} {\bibfnamefont {S.}~\bibnamefont
  {Krause}}, \bibinfo {author} {\bibfnamefont {L.}~\bibnamefont {Morrell}},
  \bibinfo {author} {\bibfnamefont {T.}~\bibnamefont {Schaerf}}, \bibinfo
  {author} {\bibfnamefont {J.}~\bibnamefont {Krause}}, \ and\ \bibinfo {author}
  {\bibfnamefont {A.}~\bibnamefont {Ward}},\ }\href@noop {} {\bibfield
  {journal} {\bibinfo  {journal} {Proc. R. Soc. B}\ }\textbf {\bibinfo {volume}
  {280}},\ \bibinfo {pages} {20122564} (\bibinfo {year} {2013})}\BibitemShut
  {NoStop}%
\bibitem [{\citenamefont {Aplin}\ \emph {et~al.}(2014)\citenamefont {Aplin},
  \citenamefont {Farine}, \citenamefont {Mann},\ and\ \citenamefont
  {Sheldon}}]{aplin2014procB}%
  \BibitemOpen
  \bibfield  {author} {\bibinfo {author} {\bibfnamefont {L.~M.}\ \bibnamefont
  {Aplin}}, \bibinfo {author} {\bibfnamefont {D.~R.}\ \bibnamefont {Farine}},
  \bibinfo {author} {\bibfnamefont {R.~P.}\ \bibnamefont {Mann}}, \ and\
  \bibinfo {author} {\bibfnamefont {B.~C.}\ \bibnamefont {Sheldon}},\
  }\href@noop {} {\bibfield  {journal} {\bibinfo  {journal} {Proc. R. Soc. B}\
  }\textbf {\bibinfo {volume} {281}},\ \bibinfo {pages} {20141016} (\bibinfo
  {year} {2014})}\BibitemShut {NoStop}%
\bibitem [{\citenamefont {Farine}\ \emph {et~al.}(2017)\citenamefont {Farine},
  \citenamefont {Strandburg-Peshkin}, \citenamefont {Couzin}, \citenamefont
  {Berger-Wolf},\ and\ \citenamefont {Crofoot}}]{farine2017procB}%
  \BibitemOpen
  \bibfield  {author} {\bibinfo {author} {\bibfnamefont {D.~R.}\ \bibnamefont
  {Farine}}, \bibinfo {author} {\bibfnamefont {A.}~\bibnamefont
  {Strandburg-Peshkin}}, \bibinfo {author} {\bibfnamefont {I.~D.}\ \bibnamefont
  {Couzin}}, \bibinfo {author} {\bibfnamefont {T.~Y.}\ \bibnamefont
  {Berger-Wolf}}, \ and\ \bibinfo {author} {\bibfnamefont {M.~C.}\ \bibnamefont
  {Crofoot}},\ }\href@noop {} {\bibfield  {journal} {\bibinfo  {journal} {Proc.
  R. Soc. B}\ }\textbf {\bibinfo {volume} {284}},\ \bibinfo {pages} {20162243}
  (\bibinfo {year} {2017})}\BibitemShut {NoStop}%
\bibitem [{\citenamefont {Couzin}\ \emph {et~al.}(2002)\citenamefont {Couzin},
  \citenamefont {Krause}, \citenamefont {James}, \citenamefont {Ruxton},\ and\
  \citenamefont {Franks}}]{couzin2002jtb}%
  \BibitemOpen
  \bibfield  {author} {\bibinfo {author} {\bibfnamefont {I.~D.}\ \bibnamefont
  {Couzin}}, \bibinfo {author} {\bibfnamefont {J.}~\bibnamefont {Krause}},
  \bibinfo {author} {\bibfnamefont {R.}~\bibnamefont {James}}, \bibinfo
  {author} {\bibfnamefont {G.~D.}\ \bibnamefont {Ruxton}}, \ and\ \bibinfo
  {author} {\bibfnamefont {N.~R.}\ \bibnamefont {Franks}},\ }\href@noop {}
  {\bibfield  {journal} {\bibinfo  {journal} {Journal of theoretical biology}\
  }\textbf {\bibinfo {volume} {218}},\ \bibinfo {pages} {1} (\bibinfo {year}
  {2002})}\BibitemShut {NoStop}%
\bibitem [{\citenamefont {Couzin}\ \emph {et~al.}(2005)\citenamefont {Couzin},
  \citenamefont {Krause}, \citenamefont {Franks},\ and\ \citenamefont
  {Levin}}]{couzin2005nature}%
  \BibitemOpen
  \bibfield  {author} {\bibinfo {author} {\bibfnamefont {I.~D.}\ \bibnamefont
  {Couzin}}, \bibinfo {author} {\bibfnamefont {J.}~\bibnamefont {Krause}},
  \bibinfo {author} {\bibfnamefont {N.~R.}\ \bibnamefont {Franks}}, \ and\
  \bibinfo {author} {\bibfnamefont {S.~A.}\ \bibnamefont {Levin}},\ }\href@noop
  {} {\bibfield  {journal} {\bibinfo  {journal} {Nature}\ }\textbf {\bibinfo
  {volume} {433}},\ \bibinfo {pages} {513} (\bibinfo {year}
  {2005})}\BibitemShut {NoStop}%
\bibitem [{\citenamefont {Jolles}\ \emph {et~al.}(2013)\citenamefont {Jolles},
  \citenamefont {King}, \citenamefont {Manica},\ and\ \citenamefont
  {Thornton}}]{jolles2013animalbeh}%
  \BibitemOpen
  \bibfield  {author} {\bibinfo {author} {\bibfnamefont {J.~W.}\ \bibnamefont
  {Jolles}}, \bibinfo {author} {\bibfnamefont {A.~J.}\ \bibnamefont {King}},
  \bibinfo {author} {\bibfnamefont {A.}~\bibnamefont {Manica}}, \ and\ \bibinfo
  {author} {\bibfnamefont {A.}~\bibnamefont {Thornton}},\ }\href@noop {}
  {\bibfield  {journal} {\bibinfo  {journal} {Animal Behaviour}\ }\textbf
  {\bibinfo {volume} {85}},\ \bibinfo {pages} {743} (\bibinfo {year}
  {2013})}\BibitemShut {NoStop}%
\bibitem [{\citenamefont {Farine}\ \emph {et~al.}(2014)\citenamefont {Farine},
  \citenamefont {Aplin}, \citenamefont {Garroway}, \citenamefont {Mann},\ and\
  \citenamefont {Sheldon}}]{farine2014animalbeh}%
  \BibitemOpen
  \bibfield  {author} {\bibinfo {author} {\bibfnamefont {D.~R.}\ \bibnamefont
  {Farine}}, \bibinfo {author} {\bibfnamefont {L.~M.}\ \bibnamefont {Aplin}},
  \bibinfo {author} {\bibfnamefont {C.~J.}\ \bibnamefont {Garroway}}, \bibinfo
  {author} {\bibfnamefont {R.~P.}\ \bibnamefont {Mann}}, \ and\ \bibinfo
  {author} {\bibfnamefont {B.~C.}\ \bibnamefont {Sheldon}},\ }\href@noop {}
  {\bibfield  {journal} {\bibinfo  {journal} {Animal behaviour}\ }\textbf
  {\bibinfo {volume} {95}},\ \bibinfo {pages} {173} (\bibinfo {year}
  {2014})}\BibitemShut {NoStop}%
\bibitem [{\citenamefont {Conradt}\ and\ \citenamefont
  {Roper}(2005)}]{conradt2005tree}%
  \BibitemOpen
  \bibfield  {author} {\bibinfo {author} {\bibfnamefont {L.}~\bibnamefont
  {Conradt}}\ and\ \bibinfo {author} {\bibfnamefont {T.~J.}\ \bibnamefont
  {Roper}},\ }\href@noop {} {\bibfield  {journal} {\bibinfo  {journal} {Trends
  in ecology \& evolution}\ }\textbf {\bibinfo {volume} {20}},\ \bibinfo
  {pages} {449} (\bibinfo {year} {2005})}\BibitemShut {NoStop}%
\bibitem [{\citenamefont {Pinkoviezky}\ \emph {et~al.}(2018)\citenamefont
  {Pinkoviezky}, \citenamefont {Couzin},\ and\ \citenamefont
  {Gov}}]{pinkoviezky2018pre}%
  \BibitemOpen
  \bibfield  {author} {\bibinfo {author} {\bibfnamefont {I.}~\bibnamefont
  {Pinkoviezky}}, \bibinfo {author} {\bibfnamefont {I.~D.}\ \bibnamefont
  {Couzin}}, \ and\ \bibinfo {author} {\bibfnamefont {N.~S.}\ \bibnamefont
  {Gov}},\ }\href@noop {} {\bibfield  {journal} {\bibinfo  {journal} {Physical
  Review E}\ }\textbf {\bibinfo {volume} {97}},\ \bibinfo {pages} {032304}
  (\bibinfo {year} {2018})}\BibitemShut {NoStop}%
\bibitem [{\citenamefont {Couzin}\ and\ \citenamefont
  {Laidre}(2009)}]{couzin2009currbiol}%
  \BibitemOpen
  \bibfield  {author} {\bibinfo {author} {\bibfnamefont {I.~D.}\ \bibnamefont
  {Couzin}}\ and\ \bibinfo {author} {\bibfnamefont {M.~E.}\ \bibnamefont
  {Laidre}},\ }\href@noop {} {\bibfield  {journal} {\bibinfo  {journal}
  {Current biology}\ }\textbf {\bibinfo {volume} {19}},\ \bibinfo {pages}
  {R633} (\bibinfo {year} {2009})}\BibitemShut {NoStop}%
\bibitem [{\citenamefont {Joshi}\ \emph {et~al.}(2017)\citenamefont {Joshi},
  \citenamefont {Couzin}, \citenamefont {Levin},\ and\ \citenamefont
  {Guttal}}]{joshi2017PLOSCompBio}%
  \BibitemOpen
  \bibfield  {author} {\bibinfo {author} {\bibfnamefont {J.}~\bibnamefont
  {Joshi}}, \bibinfo {author} {\bibfnamefont {I.~D.}\ \bibnamefont {Couzin}},
  \bibinfo {author} {\bibfnamefont {S.~A.}\ \bibnamefont {Levin}}, \ and\
  \bibinfo {author} {\bibfnamefont {V.}~\bibnamefont {Guttal}},\ }\href@noop {}
  {\bibfield  {journal} {\bibinfo  {journal} {PLoS computational biology}\
  }\textbf {\bibinfo {volume} {13}},\ \bibinfo {pages} {e1005732} (\bibinfo
  {year} {2017})}\BibitemShut {NoStop}%
\bibitem [{\citenamefont {Durham}\ and\ \citenamefont
  {Stocker}(2012)}]{durham2012AnnRewMarSci}%
  \BibitemOpen
  \bibfield  {author} {\bibinfo {author} {\bibfnamefont {W.~M.}\ \bibnamefont
  {Durham}}\ and\ \bibinfo {author} {\bibfnamefont {R.}~\bibnamefont
  {Stocker}},\ }\href@noop {} {\bibfield  {journal} {\bibinfo  {journal}
  {Annual review of marine science}\ }\textbf {\bibinfo {volume} {4}},\
  \bibinfo {pages} {177} (\bibinfo {year} {2012})}\BibitemShut {NoStop}%
\bibitem [{\citenamefont {Gueron}\ and\ \citenamefont
  {Levin}(1995)}]{gueron1995MathBioSci}%
  \BibitemOpen
  \bibfield  {author} {\bibinfo {author} {\bibfnamefont {S.}~\bibnamefont
  {Gueron}}\ and\ \bibinfo {author} {\bibfnamefont {S.~A.}\ \bibnamefont
  {Levin}},\ }\href@noop {} {\bibfield  {journal} {\bibinfo  {journal}
  {Mathematical biosciences}\ }\textbf {\bibinfo {volume} {128}},\ \bibinfo
  {pages} {243} (\bibinfo {year} {1995})}\BibitemShut {NoStop}%
\bibitem [{\citenamefont {Gueron}(1998)}]{gueron1998JMB}%
  \BibitemOpen
  \bibfield  {author} {\bibinfo {author} {\bibfnamefont {S.}~\bibnamefont
  {Gueron}},\ }\href@noop {} {\bibfield  {journal} {\bibinfo  {journal}
  {Journal of Mathematical Biology}\ }\textbf {\bibinfo {volume} {37}},\
  \bibinfo {pages} {1} (\bibinfo {year} {1998})}\BibitemShut {NoStop}%
\bibitem [{\citenamefont {Durrett}\ \emph {et~al.}(1999)\citenamefont
  {Durrett}, \citenamefont {Granovsky},\ and\ \citenamefont
  {Gueron}}]{durrett1999JourTheorProb}%
  \BibitemOpen
  \bibfield  {author} {\bibinfo {author} {\bibfnamefont {R.}~\bibnamefont
  {Durrett}}, \bibinfo {author} {\bibfnamefont {B.~L.}\ \bibnamefont
  {Granovsky}}, \ and\ \bibinfo {author} {\bibfnamefont {S.}~\bibnamefont
  {Gueron}},\ }\href@noop {} {\bibfield  {journal} {\bibinfo  {journal}
  {Journal of Theoretical Probability}\ }\textbf {\bibinfo {volume} {12}},\
  \bibinfo {pages} {447} (\bibinfo {year} {1999})}\BibitemShut {NoStop}%
\bibitem [{\citenamefont {Niwa}(2003)}]{niwa2003JTB}%
  \BibitemOpen
  \bibfield  {author} {\bibinfo {author} {\bibfnamefont {H.-S.}\ \bibnamefont
  {Niwa}},\ }\href@noop {} {\bibfield  {journal} {\bibinfo  {journal} {Journal
  of Theoretical Biology}\ }\textbf {\bibinfo {volume} {224}},\ \bibinfo
  {pages} {451} (\bibinfo {year} {2003})}\BibitemShut {NoStop}%
\bibitem [{\citenamefont {Ma}\ \emph {et~al.}(2011)\citenamefont {Ma},
  \citenamefont {Johansson},\ and\ \citenamefont {Sumpter}}]{ma2011JTB}%
  \BibitemOpen
  \bibfield  {author} {\bibinfo {author} {\bibfnamefont {Q.}~\bibnamefont
  {Ma}}, \bibinfo {author} {\bibfnamefont {A.}~\bibnamefont {Johansson}}, \
  and\ \bibinfo {author} {\bibfnamefont {D.~J.}\ \bibnamefont {Sumpter}},\
  }\href@noop {} {\bibfield  {journal} {\bibinfo  {journal} {Journal of
  theoretical biology}\ }\textbf {\bibinfo {volume} {283}},\ \bibinfo {pages}
  {35} (\bibinfo {year} {2011})}\BibitemShut {NoStop}%
\bibitem [{\citenamefont {Sueur}\ \emph {et~al.}(2011)\citenamefont {Sueur},
  \citenamefont {King}, \citenamefont {Conradt}, \citenamefont {Kerth},
  \citenamefont {Lusseau}, \citenamefont {Mettke-Hofmann}, \citenamefont
  {Schaffner}, \citenamefont {Williams}, \citenamefont {Zinner},\ and\
  \citenamefont {Aureli}}]{sueur2011oikos}%
  \BibitemOpen
  \bibfield  {author} {\bibinfo {author} {\bibfnamefont {C.}~\bibnamefont
  {Sueur}}, \bibinfo {author} {\bibfnamefont {A.~J.}\ \bibnamefont {King}},
  \bibinfo {author} {\bibfnamefont {L.}~\bibnamefont {Conradt}}, \bibinfo
  {author} {\bibfnamefont {G.}~\bibnamefont {Kerth}}, \bibinfo {author}
  {\bibfnamefont {D.}~\bibnamefont {Lusseau}}, \bibinfo {author} {\bibfnamefont
  {C.}~\bibnamefont {Mettke-Hofmann}}, \bibinfo {author} {\bibfnamefont
  {C.~M.}\ \bibnamefont {Schaffner}}, \bibinfo {author} {\bibfnamefont
  {L.}~\bibnamefont {Williams}}, \bibinfo {author} {\bibfnamefont
  {D.}~\bibnamefont {Zinner}}, \ and\ \bibinfo {author} {\bibfnamefont
  {F.}~\bibnamefont {Aureli}},\ }\href@noop {} {\bibfield  {journal} {\bibinfo
  {journal} {Oikos}\ }\textbf {\bibinfo {volume} {120}},\ \bibinfo {pages}
  {1608} (\bibinfo {year} {2011})}\BibitemShut {NoStop}%
\bibitem [{\citenamefont {Guttal}\ and\ \citenamefont
  {Couzin}(2010)}]{guttal2010pnas}%
  \BibitemOpen
  \bibfield  {author} {\bibinfo {author} {\bibfnamefont {V.}~\bibnamefont
  {Guttal}}\ and\ \bibinfo {author} {\bibfnamefont {I.~D.}\ \bibnamefont
  {Couzin}},\ }\href@noop {} {\bibfield  {journal} {\bibinfo  {journal}
  {Proceedings of the national academy of sciences}\ }\textbf {\bibinfo
  {volume} {107}},\ \bibinfo {pages} {16172} (\bibinfo {year}
  {2010})}\BibitemShut {NoStop}%
\bibitem [{\citenamefont {Torney}\ \emph {et~al.}(2010)\citenamefont {Torney},
  \citenamefont {Levin},\ and\ \citenamefont {Couzin}}]{torney2010pnas}%
  \BibitemOpen
  \bibfield  {author} {\bibinfo {author} {\bibfnamefont {C.~J.}\ \bibnamefont
  {Torney}}, \bibinfo {author} {\bibfnamefont {S.~A.}\ \bibnamefont {Levin}}, \
  and\ \bibinfo {author} {\bibfnamefont {I.~D.}\ \bibnamefont {Couzin}},\
  }\href@noop {} {\bibfield  {journal} {\bibinfo  {journal} {Proceedings of the
  National Academy of Sciences}\ }\textbf {\bibinfo {volume} {107}},\ \bibinfo
  {pages} {20394} (\bibinfo {year} {2010})}\BibitemShut {NoStop}%
\bibitem [{\citenamefont {Sridhar}\ \emph {et~al.}(2012)\citenamefont
  {Sridhar}, \citenamefont {Srinivasan}, \citenamefont {Askins}, \citenamefont
  {Canales-Delgadillo}, \citenamefont {Chen}, \citenamefont {Ewert},
  \citenamefont {Gale}, \citenamefont {Goodale}, \citenamefont {Gram},
  \citenamefont {Hart} \emph {et~al.}}]{sridhar2012amnat}%
  \BibitemOpen
  \bibfield  {author} {\bibinfo {author} {\bibfnamefont {H.}~\bibnamefont
  {Sridhar}}, \bibinfo {author} {\bibfnamefont {U.}~\bibnamefont {Srinivasan}},
  \bibinfo {author} {\bibfnamefont {R.~A.}\ \bibnamefont {Askins}}, \bibinfo
  {author} {\bibfnamefont {J.~C.}\ \bibnamefont {Canales-Delgadillo}}, \bibinfo
  {author} {\bibfnamefont {C.-C.}\ \bibnamefont {Chen}}, \bibinfo {author}
  {\bibfnamefont {D.~N.}\ \bibnamefont {Ewert}}, \bibinfo {author}
  {\bibfnamefont {G.~A.}\ \bibnamefont {Gale}}, \bibinfo {author}
  {\bibfnamefont {E.}~\bibnamefont {Goodale}}, \bibinfo {author} {\bibfnamefont
  {W.~K.}\ \bibnamefont {Gram}}, \bibinfo {author} {\bibfnamefont {P.~J.}\
  \bibnamefont {Hart}},  \emph {et~al.},\ }\href@noop {} {\bibfield  {journal}
  {\bibinfo  {journal} {The American naturalist}\ }\textbf {\bibinfo {volume}
  {180}},\ \bibinfo {pages} {777} (\bibinfo {year} {2012})}\BibitemShut
  {NoStop}%
\bibitem [{\citenamefont {Berry}\ and\ \citenamefont
  {Widder}(2014)}]{berry2014frontier}%
  \BibitemOpen
  \bibfield  {author} {\bibinfo {author} {\bibfnamefont {D.}~\bibnamefont
  {Berry}}\ and\ \bibinfo {author} {\bibfnamefont {S.}~\bibnamefont {Widder}},\
  }\href@noop {} {\bibfield  {journal} {\bibinfo  {journal} {Frontiers in
  microbiology}\ }\textbf {\bibinfo {volume} {5}} (\bibinfo {year}
  {2014})}\BibitemShut {NoStop}%
\bibitem [{\citenamefont {Graves}\ and\ \citenamefont
  {Gotelli}(1993)}]{graves1993pnas}%
  \BibitemOpen
  \bibfield  {author} {\bibinfo {author} {\bibfnamefont {G.~R.}\ \bibnamefont
  {Graves}}\ and\ \bibinfo {author} {\bibfnamefont {N.~J.}\ \bibnamefont
  {Gotelli}},\ }\href@noop {} {\bibfield  {journal} {\bibinfo  {journal}
  {Proceedings of the National Academy of Sciences}\ }\textbf {\bibinfo
  {volume} {90}},\ \bibinfo {pages} {1388} (\bibinfo {year}
  {1993})}\BibitemShut {NoStop}%
\bibitem [{\citenamefont {Ulrich}\ and\ \citenamefont
  {Gotelli}(2010)}]{ulrich2010ecology}%
  \BibitemOpen
  \bibfield  {author} {\bibinfo {author} {\bibfnamefont {W.}~\bibnamefont
  {Ulrich}}\ and\ \bibinfo {author} {\bibfnamefont {N.~J.}\ \bibnamefont
  {Gotelli}},\ }\href@noop {} {\bibfield  {journal} {\bibinfo  {journal}
  {Ecology}\ }\textbf {\bibinfo {volume} {91}},\ \bibinfo {pages} {3384}
  (\bibinfo {year} {2010})}\BibitemShut {NoStop}%
\bibitem [{\citenamefont {Sridhar}\ and\ \citenamefont
  {Shanker}(2014)}]{sridhar2014BehavEcolSocio}%
  \BibitemOpen
  \bibfield  {author} {\bibinfo {author} {\bibfnamefont {H.}~\bibnamefont
  {Sridhar}}\ and\ \bibinfo {author} {\bibfnamefont {K.}~\bibnamefont
  {Shanker}},\ }\href@noop {} {\bibfield  {journal} {\bibinfo  {journal}
  {Behavioral ecology and sociobiology}\ }\textbf {\bibinfo {volume} {68}},\
  \bibinfo {pages} {185} (\bibinfo {year} {2014})}\BibitemShut {NoStop}%
\bibitem [{\citenamefont {Majumdar}\ \emph {et~al.}(2000)\citenamefont
  {Majumdar}, \citenamefont {Krishnamurthy},\ and\ \citenamefont
  {Barma}}]{majumdar2000nonequilibrium}%
  \BibitemOpen
  \bibfield  {author} {\bibinfo {author} {\bibfnamefont {S.~N.}\ \bibnamefont
  {Majumdar}}, \bibinfo {author} {\bibfnamefont {S.}~\bibnamefont
  {Krishnamurthy}}, \ and\ \bibinfo {author} {\bibfnamefont {M.}~\bibnamefont
  {Barma}},\ }\href@noop {} {\bibfield  {journal} {\bibinfo  {journal} {Journal
  of Statistical Physics}\ }\textbf {\bibinfo {volume} {99}},\ \bibinfo {pages}
  {1} (\bibinfo {year} {2000})}\BibitemShut {NoStop}%
\bibitem [{\citenamefont {Bonabeau}\ \emph {et~al.}(1999)\citenamefont
  {Bonabeau}, \citenamefont {Dagorn},\ and\ \citenamefont
  {Fr{\'e}on}}]{bonabeau1999pnas}%
  \BibitemOpen
  \bibfield  {author} {\bibinfo {author} {\bibfnamefont {E.}~\bibnamefont
  {Bonabeau}}, \bibinfo {author} {\bibfnamefont {L.}~\bibnamefont {Dagorn}}, \
  and\ \bibinfo {author} {\bibfnamefont {P.}~\bibnamefont {Fr{\'e}on}},\
  }\href@noop {} {\bibfield  {journal} {\bibinfo  {journal} {Proceedings of the
  National Academy of Sciences}\ }\textbf {\bibinfo {volume} {96}},\ \bibinfo
  {pages} {4472} (\bibinfo {year} {1999})}\BibitemShut {NoStop}%
\bibitem [{\citenamefont {Bonabeau}\ and\ \citenamefont
  {Dagorn}(1995)}]{bonabeau1995PhysRevE}%
  \BibitemOpen
  \bibfield  {author} {\bibinfo {author} {\bibfnamefont {E.}~\bibnamefont
  {Bonabeau}}\ and\ \bibinfo {author} {\bibfnamefont {L.}~\bibnamefont
  {Dagorn}},\ }\href@noop {} {\bibfield  {journal} {\bibinfo  {journal}
  {Physical Review E}\ }\textbf {\bibinfo {volume} {51}},\ \bibinfo {pages}
  {R5220} (\bibinfo {year} {1995})}\BibitemShut {NoStop}%
\bibitem [{\citenamefont {Griesser}\ \emph {et~al.}(2011)\citenamefont
  {Griesser}, \citenamefont {Ma}, \citenamefont {Webber}, \citenamefont
  {Bowgen},\ and\ \citenamefont {Sumpter}}]{griesser2011PLOSone}%
  \BibitemOpen
  \bibfield  {author} {\bibinfo {author} {\bibfnamefont {M.}~\bibnamefont
  {Griesser}}, \bibinfo {author} {\bibfnamefont {Q.}~\bibnamefont {Ma}},
  \bibinfo {author} {\bibfnamefont {S.}~\bibnamefont {Webber}}, \bibinfo
  {author} {\bibfnamefont {K.}~\bibnamefont {Bowgen}}, \ and\ \bibinfo {author}
  {\bibfnamefont {D.~J.}\ \bibnamefont {Sumpter}},\ }\href@noop {} {\bibfield
  {journal} {\bibinfo  {journal} {PLoS One}\ }\textbf {\bibinfo {volume} {6}},\
  \bibinfo {pages} {e23438} (\bibinfo {year} {2011})}\BibitemShut {NoStop}%
\bibitem [{\citenamefont {Strogatz}(1994)}]{strogatz2014book}%
  \BibitemOpen
  \bibfield  {author} {\bibinfo {author} {\bibfnamefont {S.~H.}\ \bibnamefont
  {Strogatz}},\ }\href@noop {} {\emph {\bibinfo {title} {Nonlinear dynamics and
  chaos: with applications to physics, biology, chemistry, and engineering}}}\
  (\bibinfo  {publisher} {Westview press},\ \bibinfo {year} {1994})\BibitemShut
  {NoStop}%
\bibitem [{\citenamefont {Van~Kampen}(1992)}]{van1992stochastic}%
  \BibitemOpen
  \bibfield  {author} {\bibinfo {author} {\bibfnamefont {N.~G.}\ \bibnamefont
  {Van~Kampen}},\ }\href@noop {} {\emph {\bibinfo {title} {Stochastic processes
  in physics and chemistry}}},\ Vol.~\bibinfo {volume} {1}\ (\bibinfo
  {publisher} {Elsevier},\ \bibinfo {year} {1992})\BibitemShut {NoStop}%
\bibitem [{\citenamefont {Horsthemke}\ and\ \citenamefont
  {Lefever}(1984)}]{horsthemke1984noise}%
  \BibitemOpen
  \bibfield  {author} {\bibinfo {author} {\bibfnamefont {W.}~\bibnamefont
  {Horsthemke}}\ and\ \bibinfo {author} {\bibfnamefont {R.}~\bibnamefont
  {Lefever}},\ }\href@noop {} {\bibfield  {journal} {\bibinfo  {journal}
  {Noise-induced transitions: theory and applications in physics, chemistry,
  and biology}\ ,\ \bibinfo {pages} {164}} (\bibinfo {year}
  {1984})}\BibitemShut {NoStop}%
\bibitem [{\citenamefont {Biancalani}\ \emph {et~al.}(2014)\citenamefont
  {Biancalani}, \citenamefont {Dyson},\ and\ \citenamefont
  {McKane}}]{biancalani2014noise}%
  \BibitemOpen
  \bibfield  {author} {\bibinfo {author} {\bibfnamefont {T.}~\bibnamefont
  {Biancalani}}, \bibinfo {author} {\bibfnamefont {L.}~\bibnamefont {Dyson}}, \
  and\ \bibinfo {author} {\bibfnamefont {A.~J.}\ \bibnamefont {McKane}},\
  }\href@noop {} {\bibfield  {journal} {\bibinfo  {journal} {Physical review
  letters}\ }\textbf {\bibinfo {volume} {112}},\ \bibinfo {pages} {038101}
  (\bibinfo {year} {2014})}\BibitemShut {NoStop}%
\bibitem [{\citenamefont {Jhawar}\ \emph {et~al.}(2019)\citenamefont {Jhawar},
  \citenamefont {Morris},\ and\ \citenamefont {Guttal}}]{jhawar2018deriving}%
  \BibitemOpen
  \bibfield  {author} {\bibinfo {author} {\bibfnamefont {J.}~\bibnamefont
  {Jhawar}}, \bibinfo {author} {\bibfnamefont {R.~G.}\ \bibnamefont {Morris}},
  \ and\ \bibinfo {author} {\bibfnamefont {V.}~\bibnamefont {Guttal}},\ }\href
  {\doibase https://doi.org/10.1016/bs.host.2018.10.002} {\emph {\bibinfo
  {title} {Integrated Population Biology and Modeling, Part B}}},\ \bibinfo
  {series} {Handbook of Statistics}, Vol.~\bibinfo {volume} {40}\ (\bibinfo
  {publisher} {Elsevier},\ \bibinfo {year} {2019})\ pp.\ \bibinfo {pages} {551
  -- 594}\BibitemShut {NoStop}%
\bibitem [{\citenamefont {Ward}\ \emph {et~al.}(2008)\citenamefont {Ward},
  \citenamefont {Sumpter}, \citenamefont {Couzin}, \citenamefont {Hart},\ and\
  \citenamefont {Krause}}]{ward2008quorum}%
  \BibitemOpen
  \bibfield  {author} {\bibinfo {author} {\bibfnamefont {A.~J.}\ \bibnamefont
  {Ward}}, \bibinfo {author} {\bibfnamefont {D.~J.}\ \bibnamefont {Sumpter}},
  \bibinfo {author} {\bibfnamefont {I.~D.}\ \bibnamefont {Couzin}}, \bibinfo
  {author} {\bibfnamefont {P.~J.}\ \bibnamefont {Hart}}, \ and\ \bibinfo
  {author} {\bibfnamefont {J.}~\bibnamefont {Krause}},\ }\href@noop {}
  {\bibfield  {journal} {\bibinfo  {journal} {Proceedings of the National
  Academy of Sciences}\ }\textbf {\bibinfo {volume} {105}},\ \bibinfo {pages}
  {6948} (\bibinfo {year} {2008})}\BibitemShut {NoStop}%
\end{thebibliography}%

\onecolumngrid
\pagebreak
%\vspace{2in}
\section*{Figures}

\begin{figure}[H]
     \begin{center}
     \begin{tabular}{ M{3cm}  M{3cm}  M{0.6\textwidth}  }
     \hline
        Event Description & Change in Number of Focal Groups/Rate & Transition Event\\ 
    \hline
      Two smaller groups merge to create a new focal group
      & 
      +1/$Q_{+1}$
      &
      \raisebox{-\totalheight}{\includegraphics[width=0.6\textwidth]{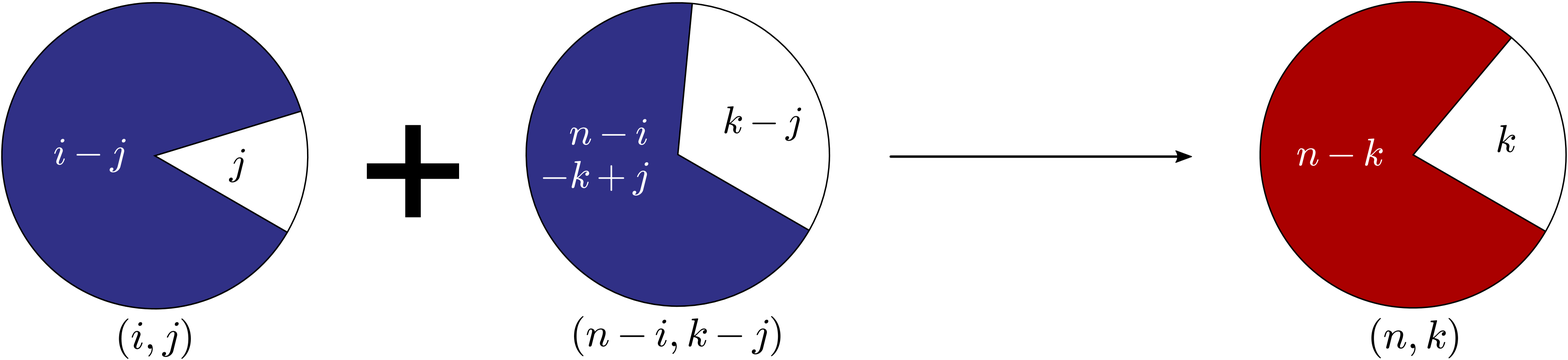}}
      \\ \hline
      
      Focal group merges with a group of different composition
      & 
      -1/$Q_{-1}$
      & 
      \raisebox{-\totalheight}{\includegraphics[width=0.6\textwidth]{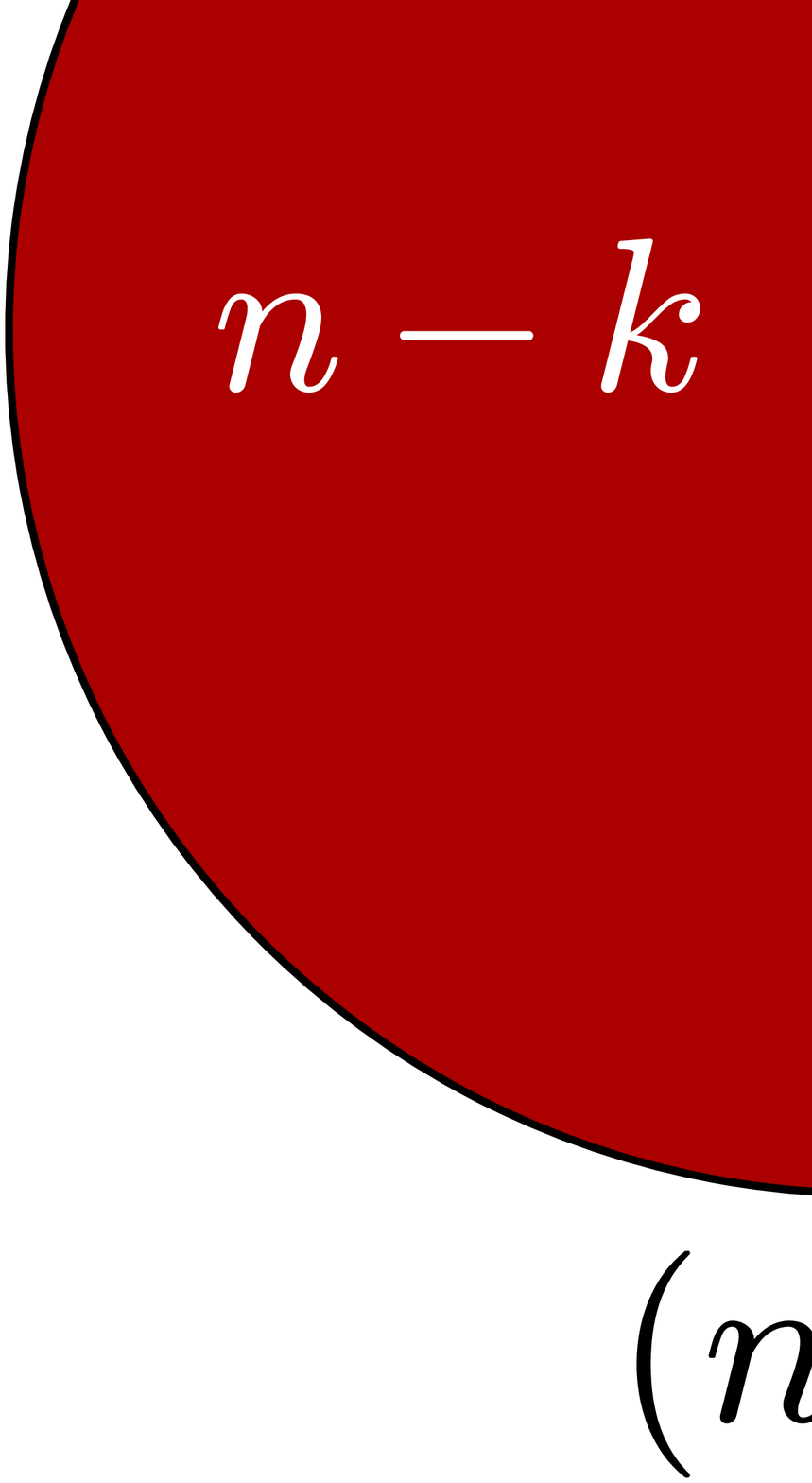}}
      \\ \hline
      
      Two focal groups merge
      & 
      -2/$Q_{-2}$
      &
      \raisebox{-\totalheight}{\includegraphics[width=0.6\textwidth]{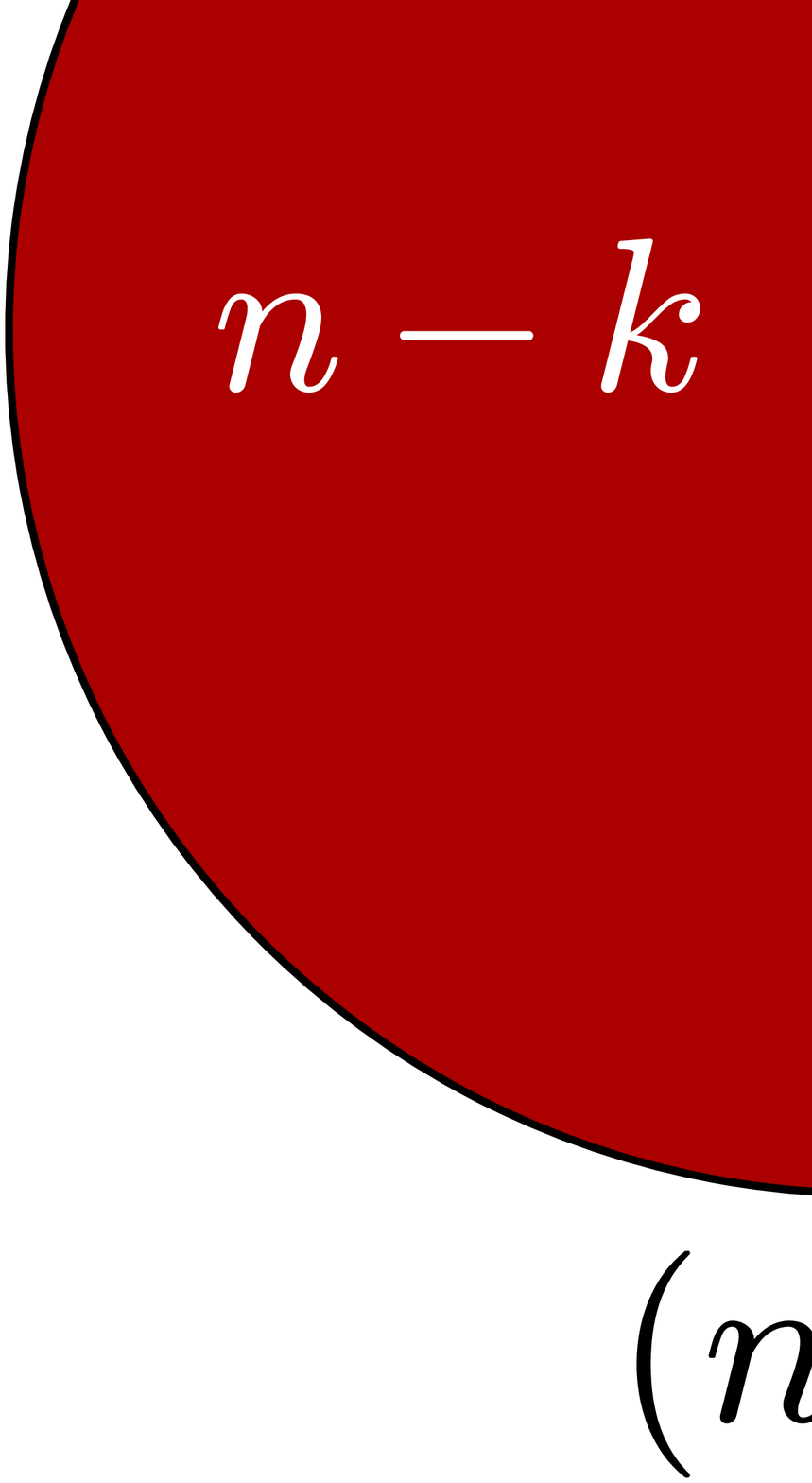}}
      \\ \hline
      
      Larger group splits to create a new focal group
      & 
      +1/$P_{+1}$
      & 
      \raisebox{-\totalheight}{\includegraphics[width=0.6\textwidth]{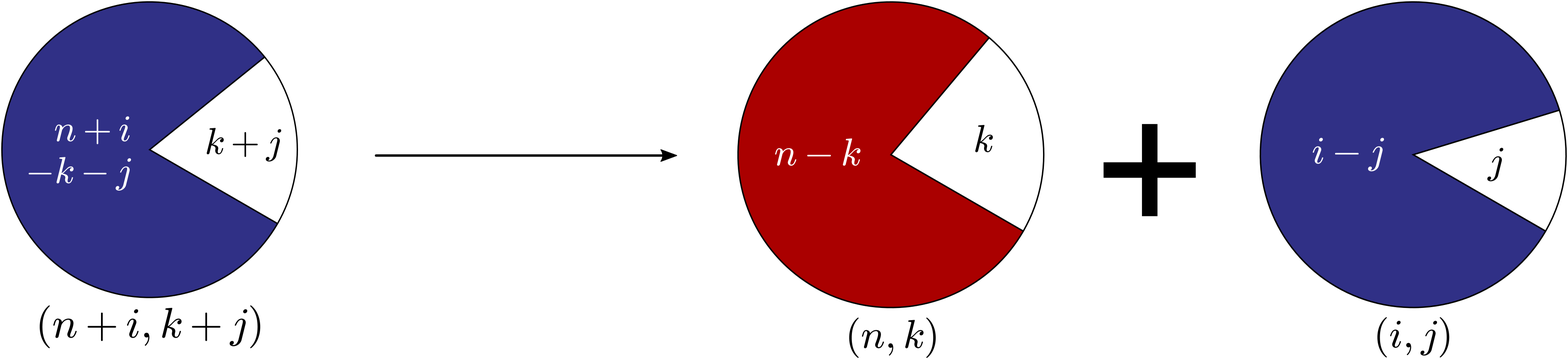}}
      \\ \hline
      
      Group of twice the size and composition splits to form two focal groups
      & 
      +2/$P_{+2}$
      &
      \raisebox{-\totalheight}{\includegraphics[width=0.6\textwidth]{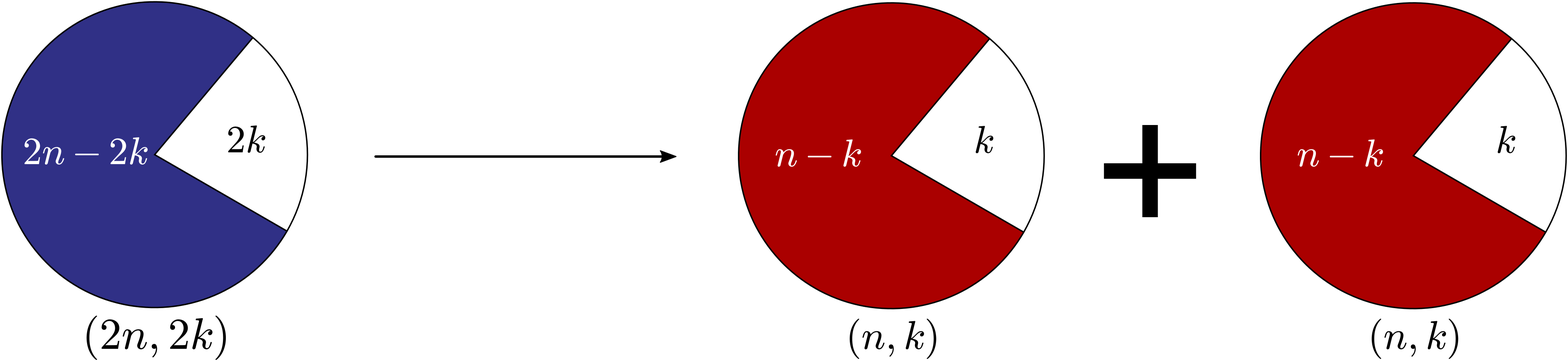}}
      \\ \hline
      
      Focal group splits into smaller groups
      & 
      -1/$P_{-1}$
      &
      \raisebox{-\totalheight}{\includegraphics[width=0.6\textwidth]{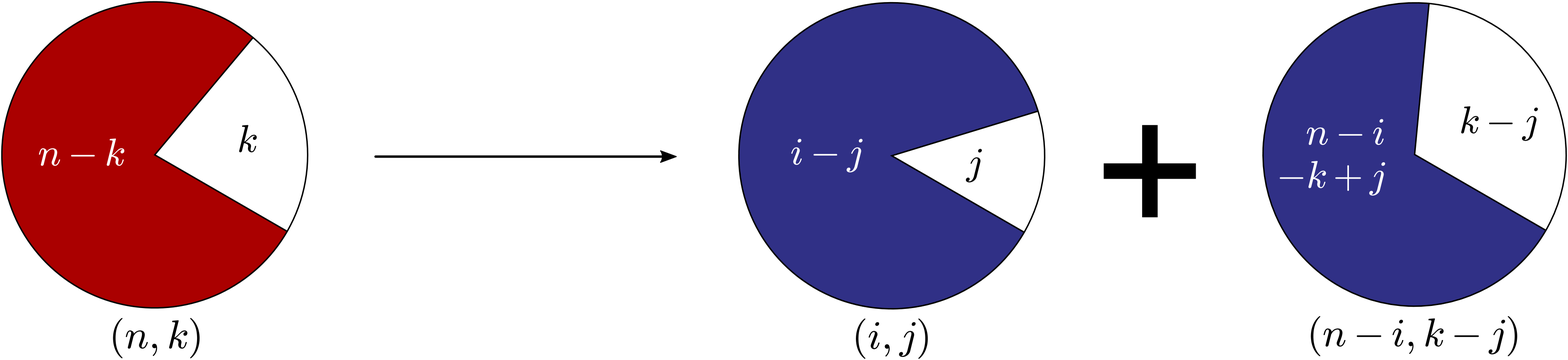}}
      \\ \hline
      & & \multicolumn{1}{r}{\includegraphics[width=2in]{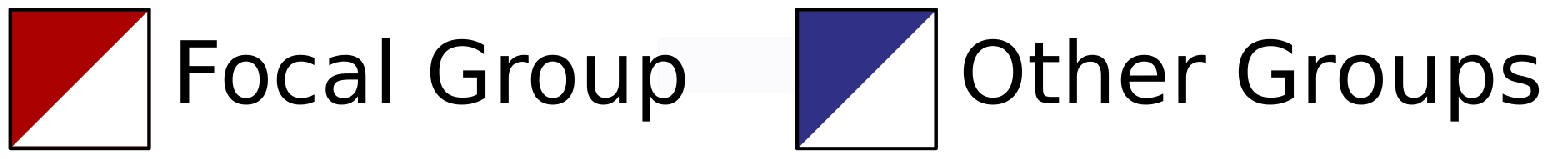}}
      \\
      \end{tabular}
      \caption{All the possible transition events are presented here graphically. Each pie-chart represents a group with the white region corresponding to type-I and the coloured region (blue or red) to type-II. The two regions have been labelled with the number of type-I and type-II individuals in each group. Underneath every group we label it with $(m,l)$ where $m$ is the size of the group and $l$ is the number of type-I individuals in the group. We focus our attention on $(n,k)$ groups, which use the red-white colour scheme, while other groups are blue-white. The notation for rates, $P_\alpha$ and $Q_\alpha$ in the second column are defined in~\ref{subsec:transition}. Note that the above figures are purely representational; in this model, we do not consider  geometry of groups or how the two types of species are structured within groups.}
      \label{fig:transitions}
      \end{center}
      \end{figure}



\end{document}